\documentclass[twoside,twocolumn,9pt]{article}
\usepackage{extsizes}
\usepackage[super,sort&compress,comma]{natbib} 
\usepackage[version=3]{mhchem}
\usepackage[left=1.5cm, right=1.5cm, top=1.785cm, bottom=2.0cm]{geometry}
\usepackage{balance}
\usepackage{amssymb}
\usepackage{amsthm}
\usepackage{amsmath,bm}
\usepackage{mathptmx}
\usepackage{sectsty}
\usepackage{graphicx} 
\usepackage{lastpage}
\usepackage[format=plain,justification=justified,singlelinecheck=false,font={stretch=1.125,small,sf},labelfont=bf,labelsep=space]{caption}
\usepackage{float}
\usepackage{fancyhdr}
\usepackage{fnpos}
\usepackage[english]{babel}
\usepackage{array}
\usepackage{droidsans}
\usepackage{charter}
\usepackage[T1]{fontenc}
\usepackage[usenames,dvipsnames]{xcolor}
\usepackage{setspace}
\usepackage[compact]{titlesec}
\usepackage{hyperref}

\usepackage{epstopdf}

\definecolor{cream}{RGB}{222,217,201}

\usepackage{caption}
\usepackage{subcaption}
\usepackage{hyperref} 
\usepackage{cuted} 
\usepackage{graphicx}  
\usepackage{arydshln}
\usepackage{amsmath,amsfonts,amssymb}
\DeclareMathOperator*{\argmax}{arg\,max}

\definecolor{color_Shu1}{rgb}{0.3 0.7 1}
\definecolor{color_Shu2}{rgb}{1 0 1}
\usepackage[normalem]{ulem}
\usepackage{tabu} 
\usepackage{float} 
\usepackage[utf8]{inputenc}
\usepackage{booktabs, caption, makecell}

\usepackage{threeparttable}
\usepackage{xcolor}
\usepackage{algorithm}
\usepackage{algpseudocode}
\usepackage{multirow} 
\usepackage{tikz}
\usetikzlibrary{decorations.pathmorphing} 
\usetikzlibrary{fit}					
\usetikzlibrary{backgrounds}	
\usepackage{makecell}

\algdef{SE}[DOWHILE]{Do}{doWhile}{\algorithmicdo}[1]{\algorithmicwhile\ #1}

\begin{document}

\pagestyle{fancy}
\thispagestyle{plain}
\fancypagestyle{plain}{
\renewcommand{\headrulewidth}{0pt}
}

\makeFNbottom
\makeatletter
\renewcommand\LARGE{\@setfontsize\LARGE{15pt}{17}}
\renewcommand\Large{\@setfontsize\Large{12pt}{14}}
\renewcommand\large{\@setfontsize\large{10pt}{12}}
\renewcommand\footnotesize{\@setfontsize\footnotesize{7pt}{10}}
\makeatother

\renewcommand{\thefootnote}{\fnsymbol{footnote}}
\renewcommand\footnoterule{\vspace*{1pt}%
\color{cream}\hrule width 3.5in height 0.4pt \color{black}\vspace*{5pt}} 
\setcounter{secnumdepth}{5}

\makeatletter 
\renewcommand\@biblabel[1]{#1}            
\renewcommand\@makefntext[1]%
{\noindent\makebox[0pt][r]{\@thefnmark\,}#1}
\makeatother 
\renewcommand{\figurename}{\small{Fig.}~}
\sectionfont{\sffamily\Large}
\subsectionfont{\normalsize}
\subsubsectionfont{\bf}
\setstretch{1.125} 
\setlength{\skip\footins}{0.8cm}
\setlength{\footnotesep}{0.25cm}
\setlength{\jot}{10pt}
\titlespacing*{\section}{0pt}{4pt}{4pt}
\titlespacing*{\subsection}{0pt}{15pt}{1pt}

\fancyfoot{}
\fancyfoot[LO,RE]{\vspace{-7.1pt}\includegraphics[height=9pt]{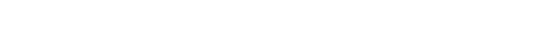}}
\fancyfoot[CO]{\vspace{-7.1pt}\hspace{13.2cm}\includegraphics{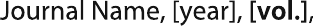}}
\fancyfoot[CE]{\vspace{-7.2pt}\hspace{-14.2cm}\includegraphics{head_foot/RF}}
\fancyfoot[RO]{\footnotesize{\sffamily{1--\pageref{LastPage} ~\textbar  \hspace{2pt}\thepage}}}
\fancyfoot[LE]{\footnotesize{\sffamily{\thepage~\textbar\hspace{3.45cm} 1--\pageref{LastPage}}}}
\fancyhead{}
\renewcommand{\headrulewidth}{0pt} 
\renewcommand{\footrulewidth}{0pt}
\setlength{\arrayrulewidth}{1pt}
\setlength{\columnsep}{6.5mm}
\setlength\bibsep{1pt}

\makeatletter 
\newlength{\figrulesep} 
\setlength{\figrulesep}{0.5\textfloatsep} 

\newcommand{\topfigrule}{\vspace*{-1pt}%
\noindent{\color{cream}\rule[-\figrulesep]{\columnwidth}{1.5pt}} }

\newcommand{\botfigrule}{\vspace*{-2pt}%
\noindent{\color{cream}\rule[\figrulesep]{\columnwidth}{1.5pt}} }

\newcommand{\dblfigrule}{\vspace*{-1pt}%
\noindent{\color{cream}\rule[-\figrulesep]{\textwidth}{1.5pt}} }

\makeatother

\twocolumn[
  \begin{@twocolumnfalse}
{\includegraphics[height=30pt]{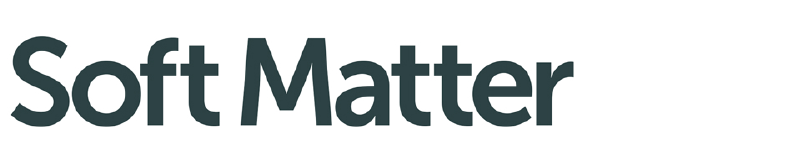}\hfill\raisebox{0pt}[0pt][0pt]{\includegraphics[height=55pt]{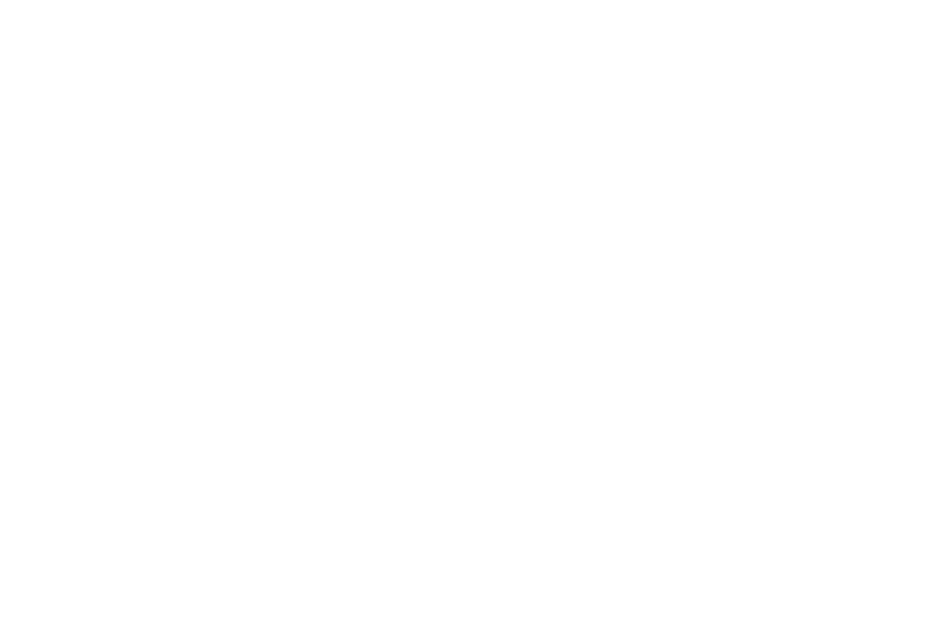}}\\[1ex]
\includegraphics[width=18.5cm]{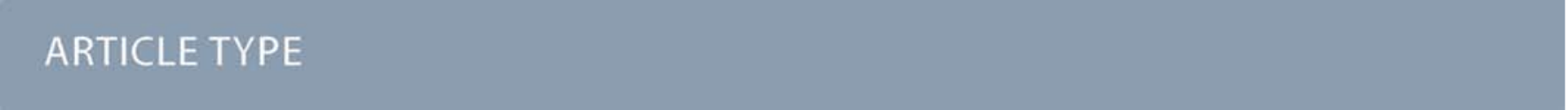}}\par
\vspace{1em}
\sffamily
\begin{tabular}{m{4.5cm} p{13.5cm} }

\includegraphics{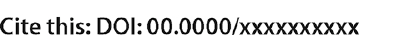} &
\noindent\LARGE{\textbf{Transfer Learning of Memory Kernels in Coarse-grained Modeling}} \\
\vspace{0.3cm} & \vspace{0.3cm} \\

& \noindent\large{Zhan Ma,\textit{$^{a}$} Shu Wang,\textit{$^{a}$} 
Minhee Kim,\textit{$^{b}$}
Kaibo Liu,\textit{$^{b}$}
Chun-Long Chen,\textit{$^{c}$}
and Wenxiao Pan\textit{$^{a}$}$^{\ast}$} \\

\includegraphics{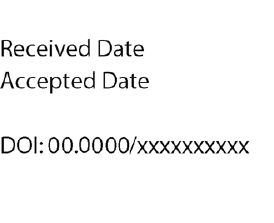} & 
\noindent\normalsize{
The present work concerns the transferability of coarse-grained (CG) modeling in reproducing the dynamic properties of the reference atomistic systems across a range of parameters. In particular, we focus on implicit-solvent CG modeling of polymer solutions. The CG model is based on the generalized Langevin equation, where the memory kernel plays the critical role in determining the dynamics in all time scales. Thus, we propose methods for transfer learning of memory kernels. The key ingredient of our methods is Gaussian process regression. By integration with the model order reduction via proper orthogonal decomposition and the active learning technique, the transfer learning can be practically efficient and requires minimum training data. Through two example polymer solution systems, we demonstrate the accuracy and efficiency of the proposed transfer learning methods in the construction of transferable memory kernels. The transferability allows for out-of-sample predictions, even in the extrapolated domain of parameters. Built on the transferable memory kernels, the CG models can reproduce the dynamic properties of polymers in all time scales at different thermodynamic conditions (such as temperature and solvent viscosity) and for different systems with varying concentrations and lengths of polymers.

} 

\end{tabular}

 \end{@twocolumnfalse} \vspace{0.6cm}

]

\renewcommand*\rmdefault{bch}\normalfont\upshape
\rmfamily
\section*{}
\vspace{-1cm}
\footnotetext{\textit{$^{a}$~Department of Mechanical Engineering, University of Wisconsin-Madison, Madison, WI 53706. }}
\footnotetext{\textit{$^{b}$~Department of Industrial and Systems Engineering, University of Wisconsin-Madison, Madison, WI 53706. }}
\footnotetext{\textit{$^{c}$~Physical Sciences Division, Pacific Northwest National Laboratory, Richland, WA 99352.}}
\footnotetext{$^{\ast}$~Corresponding author. E-mail: wpan9@wisc.edu}





\section{Introduction}
\label{sec:Introduction}
To study polymers or biomolecules in solution, coarse-grained (CG) modeling and simulations can be practically more efficient \cite{saunders2013coarse,kmiecik2016coarse,dinpajooh2018coarse,salerno2016resolving, gooneie2017review,CG_peptoid1_2020,CG_peptoid2_2016,CG_peptoid3_2018}, compared with full atomistic simulations via, e.g., all-atom molecular dynamics (MD).
Instead of tracking individual atoms of molecules and solvent, the CG modeling averages out or eliminates certain degrees of freedom (DOFs) to reduce the system's dimensionality and captures the molecules' collective dynamics and properties. The removal of highly-fluctuating atomic DOFs and the larger characteristic length scale of CG coordinates permit to employ larger time steps in CG simulations. In implicit-solvent CG modeling \cite{kleinjung2014design,pham2008brownian,chudoba2018tuning,sevink2014efficient,Pan_CGMZ_2019,Pan_DCGSD_2020}, not only the DOFs representing polymer molecules are reduced, but also the solvent DOFs are eliminated. Provided significantly reduced DOFs and larger time steps, CG modeling is computationally more efficient than full atomistic simulations, and hence, can grant larger accessible length scales and render tractable simulating long-time effects in practical applications \cite{mills2013mesoscale,mu2016comparative,rovigatti2019numerical,beldowski2018physical}. 

However, to reap the benefit of CG modeling, two challenges must be addressed. The first one is to conserve both the structural and dynamic properties of polymers under coarse-graining. To conserve the structural properties (e.g., radial and angular distribution functions), the CG potential (or potential of mean force) must be correctly constructed \cite{reith2003deriving,lyubartsev1995calculation,izvekov2005multiscale,shell2008relative,sanyal2016coarse,GaussianCGPotential_2017,zhang2018deepcg,ML_CGMD_Wang2019}. To conserve the dynamic properties (e.g., diffusivity and velocity autocorrelation function (VACF)), the kinetic effect of unresolved DOFs (including solvent) on the system must be properly accounted in CG modeling. For that, a non-Markovian dynamics (e.g., in the form of generalized Langevin equation (GLE)) must be introduced in the CG model because the elimination of DOFs results in a non-Markovian memory in the dynamics of CG variables, as discussed in literature \cite{li2017computing,ma2016derivation,lei2016data,jung2017iterative,jung2018generalized,lee2019multi} and also in our prior work \cite{Pan_CGMZ_2019,Pan_DCGSD_2020}. The second challenge lies in the transferability of CG modeling, for instance, how the CG model constructed can be transferable across different thermodynamic conditions. The efforts to attain transferable CG potentials that preserve structural properties under coarse-graining have made substantial progress \cite{CGGNN_JCP2020,CGGNN_Ionic_JCP2020,CGTrans_JCTC2020,CGTrans_JCTC2017,CGTrans_JCP2020}. In contrast, the transferability of a CG model in conserving the dynamic properties has not been extensively discussed. Lyubimov et al. derived from the GLE an analytical factor for dynamical rescaling of the friction coefficient in CG modeling to correctly capture the long-time diffusion of polymers. The derived rescaling factor is transferable for different polymer systems and thermodynamic conditions, with the temperature and radius-of-gyration as the input parameters \cite{AnalyticalRescaling_Lyubimov2010,AnalyticalRescaling_Lyubimov2011}. However, the approach is only applicable to polymer melts and reproducing the long-time diffusion coefficient of polymers \cite{AnalyticalRescaling_Lyubimov2010,AnalyticalRescaling_Lyubimov2011}. For polymers in solution and to reproduce the dynamic properties beyond the long-time normal diffusion, e.g., super- and/or sub-diffusion, and the VACF as a function of time, it calls for new methodology to render transferability in CG modeling. Noting that the memory kernel plays a critical role for a CG model to reproduce the entire dynamics \cite{Bian2016,li2017computing,ma2016derivation,yoshimoto2017construction}, especially in implicit-solvent CG modeling with many solvent DOFs unresolved~\cite{lei2016data,jung2017iterative,Pan_CGMZ_2019,Pan_DCGSD_2020}, the present work hence focuses on the transferability of the memory kernel in CG modeling. 

In particular, we propose two transfer learning methods to enable the memory kernel transferable across different thermodynamic conditions (such as temperature and solvent viscosity) and across different systems with varying solute concentrations and lengths of polymer chains. The proposed transfer learning methods draw on Gaussian process regression (GPR). By integration with the model order reduction via proper orthogonal decomposition (POD) and the active learning technique, the transfer learning can be practically efficient and requires minimum training data. GPR is chosen because of its flexibility in both interpolation and extrapolation and its capability for prediction with quantified uncertainty. The model order reduction enables to represent the memory kernel (as a function of time and parameters) in a reduced temporal and parameter space, which in turn greatly reduces the training and prediction costs of GPR. The active learning technique allows for efficient use of data with maximum information gain via adaptive sampling guided by the uncertainty quantified in GPR. Through two example polymer solution systems, we demonstrate the accuracy and efficiency of the proposed transfer learning methods in the construction of transferable memory kernels, from which the CG models can reproduce the dynamic properties of polymers across different thermodynamic conditions and systems.
 
The rest of the paper is organized as follows. In \S\ref{subsec:CG_modeling}, we describe the framework of CG modeling, which is built upon the GLE and extended dynamics. \S\ref{subsec:transfer_learning} explains in detail the proposed transfer learning methods, whose key ingredients include GPR, POD, and active learning. We present the numerical results in \S\ref{sec:results}, where two benchmark examples are used to assess the accuracy and computational cost of transfer learning. Finally, we conclude and summarize our main findings and contributions in \S\ref{sec:conclusion}.
\section{Methodology} \label{sec:method}
\subsection{CG modeling}\label{subsec:CG_modeling}
Without loss of generality, we consider an atomistic system consisting of $n$ atoms in polymer molecules, with coordinates $\mathbf{r} = \{ \mathbf{r}_i | i=1, 2, \dots, n \}$ and momenta $\mathbf{p} = \{\mathbf{p}_i | i=1, 2, \dots, n\}$. In CG modeling, $n$ atoms are coarse-grained as $N$ clusters (referred to as CG particles), and each cluster contains $n_c$ atoms. 
To be consistent in notation, we use the lowercase $m_i$, $\mathbf{r}_i$, and $\mathbf{p}_i$ to represent the mass, position, and momentum of the $i$-th atom in the atomistic system; and the uppercase $M_I$, $\mathbf{R}_I$, and $\mathbf{P}_I$ denote the mass, position, and momentum of the $I$-th CG particle in the CG system. The variables of the atomistic and CG systems are related via:
\begin{equation}\label{Eq:RI_PI}
M_I = \sum\limits_{i=1}^{n_c} m_{Ii}\;, ~~~~\mathbf{R}_I = \frac{1}{M_I}\sum\limits_{i=1}^{n_c} m_{Ii} \mathbf{r}_{Ii}\;, ~~~~ \mathbf{P}_I = \sum\limits_{i=1}^{n_c} \mathbf{p}_{Ii} \; ,
\end{equation}
where the subscript ${Ii}$ denotes the $i$-th atom in the $I$-th CG particle; $M_I$, $\mathbf{R}_I$, and $\mathbf{P}_I$ are defined as the total mass, center-of-mass (COM) position, and total momentum of all atoms in the $I$-th CG particle, respectively.  

The GLE given by the Mori-Zwanzig projection formalism \cite{mori1965transport,zwanzig1973r,zwanzig2001nonequilibrium} provides a theoretically sound framework for CG modeling, where the non-Markovian dynamics of the CG system is governed by:    
\begin{equation}
\label{Eq:gle}
\dot{\mathbf{P}}_I = \langle \mathbf{F}_I \rangle - \int_{0}^{t} K (t-t')
 M_I^{-1}\mathbf{P}_{I}(t') dt'+\tilde {\mathbf{F}}_{I}  \; .
\end{equation}
In the right-hand side of Eq.~\eqref{Eq:gle}, the first term represents the mean force on the $I$-th CG particle. The third term $\tilde{\mathbf{F}}_I$ denotes the random force. The second term (referred to as the dissipative force) has a memory kernel $K (t-t')$, which is related to the random force by $K(t)=(1/k_BT) \langle [\tilde {\mathbf{F}}_{I}(t)]^\intercal[\tilde {\mathbf{F}}_I(0)]\rangle$ to satisfy the second fluctuation-dissipation theorem \cite{kubo1966r}. Here, $k_B$ denotes the Boltzmann constant; $T$ is the thermodynamic temperature. The kinetic effect of the lost atomic DOFs under coarse-graining is properly accounted in the CG dynamics by the memory kernel and the random force with colored noise in Eq.~\eqref{Eq:gle} \cite{zwanzig1973r,zwanzig2001nonequilibrium}.

The mean force in Eq.~\eqref{Eq:gle} is $\langle \mathbf{F}_I \rangle = \frac{1}{k_B T}\frac{\partial }{\partial \mathbf{R}_{I}} \ln\omega(\mathbf{R}) $, where $\mathbf{R} = \{\mathbf{R}_1, \mathbf{R}_2, \dots ,\mathbf{R}_N\} $ is a point in the CG phase space; $\omega(\mathbf{R})$ represents a normalized partition function of all the atomistic configurations at phase point $\mathbf{R}$. Since the present work concerns the dynamic properties and does not consider the structural properties or free energy, $\langle F_I\rangle$ is regarded as the average for the $I$-th CG particle over all phase points. Thus, without external force fields, the mean force exerted on a CG particle is approximated to be zero; i.e., $\langle \mathbf{F}_I \rangle = 0$. Eq.~\eqref{Eq:gle} can hence be simplified to:
\begin{equation}\label{Eq:gle2}
\dot{\mathbf{P}}_{I}(t) = - \int_{0}^{t}  {K}(t-t') \mathbf{V}_I (t')dt'+\tilde{ \mathbf{F}}_{I}(t) \; ,
\end{equation}
with the velocity $\mathbf{V}_I(t') = \frac{\mathbf{P}_I(t')}{M_I}$. 
 
To conserve the dynamics and reproduce the dynamic properties of the underlying atomistic system, the memory kernel in Eq.~\eqref{Eq:gle2} must be directly linked to the atomistic system and can be computed from the atomistic data.

\subsubsection{Memory kernel}\label{subsubsec:memory}
To determine the memory kernel ${K}(t)$ in Eq.~\eqref{Eq:gle2} from the atomistic data, we rely on the property that the velocity $\mathbf{V}$ and random force $\tilde{\mathbf{F}}$ come from two orthogonal subspaces and hence are not correlated to each other, i.e., $\langle \mathbf{V}^T \tilde{\mathbf{F}}  \rangle = 0$. Thus, multiplying both sides of Eq.~\eqref{Eq:gle2} by $\mathbf{V}(0)^T$ leads to:
\begin{equation}\label{Eq:gle3}
\langle \mathbf{V}(0)^T \dot{ \mathbf{P}}(t) \rangle = - \int_{0}^{t}  {K}(t-t') \langle \mathbf{V}(0)^T \mathbf{V}(t') \rangle dt' \; ,
\end{equation}
where $C(t)= \langle \mathbf{V}(0)^T \mathbf{V}(t') \rangle $ is the VACF; $W(t)=\langle \mathbf{V}(0)^T \dot{\mathbf{P}}(t) \rangle$ defines the force-velocity correlation function (FVCF). While the data of VACF can be directly obtained in the atomistic simulations, the FVCF can be evaluated from $W(t)=\frac{d C(t)}{d t}$ using numerical differentiation. The memory kernel ${K}(t)$ can then be solved via deconvolution of Eq.~\eqref{Eq:gle3} given the data of VACF and FVCF. Note that the integral operator in Eq.~\eqref{Eq:gle3} is the Volterra operator \cite{Linz1985Volterra}. Although the deconvolution can have a unique continuous solution, the solution does not depend continuously on the data; i.e., the solution is unstable against data noise due to the nondegeneracy of $C(t)$. Thus, solving the deconvolution to determine the memory kernel is an ill-posed problem, and proper regularization must be enforced. What we have are the data of VACF and FVCF at discrete times (i.e., $t_i = (i-1) \Delta t$ with $i = 1,2,\dots, N_t$). Hence, Eq.~\eqref{Eq:gle3} can be discretized, and the deconvolution is solved in the discrete setting. Note that discretization can regularize an ill-posed problem, known as the ``self-regularization" property of discretization \cite{Lamm2000Volterra_survey}. As a result, the deconvolution problem becomes well-posed when solved in the discrete setting. However, the linear system resulting from discretization could be ill-conditioned \cite{Groetsch2015Linear_inverse} and require further regularization. Applying the midpoint quadrature rule \cite{Linz1985Volterra}, Eq.~\eqref{Eq:gle3} can be discretized into the following linear system:
\begin{equation}\label{Eq:discrete_decon}
    \mathbf{C} \mathbf{K} = -\mathbf{W} \;,
\end{equation}
where $\mathbf{C} \in \mathbb{R}^{(N_t-1) \times (N_t-1)} $ with
\begin{equation*}\label{Eq:C_matrix}
\mathbf{C}_{i,j} =
    \begin{cases}
      \frac{\Delta t }{2}(C(t_{i-j+1}) + C(t_{i-j+2})) & i \geq j \\
      0 & i < j\\
    \end{cases} \; ;  
\end{equation*}
$\mathbf{K} \in \mathbb{R}^{N_t-1} $ with $\mathbf{K}_i = K(t_{i+1/2}) $; and $\mathbf{W} \in \mathbb{R}^{N_t-1} $ with $\mathbf{W}_i = W(t_{i+1}) $. If the linear system in Eq.~\eqref{Eq:discrete_decon} is ill-conditioned, which means its solution is unstable and sensitive to data noise, the Tikhonov regularization \cite{Tikhonov1963Tikhonov1,Tikhonov1963Tikhonov2} is introduced and leads to the following regularized linear system:
\begin{equation}\label{Eq:discrete_decon_reg}
    (\mathbf{C}^T \mathbf{C} +\beta)\mathbf{K} = - \mathbf{C}^T \mathbf{W} \;,
\end{equation}
where $\mathbf{C}^T $ is the transpose of $\mathbf{C}$, and $\beta $ is the regularization parameter. The value of $\beta $ can be determined using the quasi-optimality criterion \cite{tikhonov1964approximate,tikhonov1965use}. In addition to the deconvolution, the numerical differentiation used to obtain the data of FVCF is also an ill-posed problem and requires regularization. 

For the numerical examples considered in this work,  the FVCF ($W(t)$) was obtained via numerical differentiation regularized by the Tikhonov regularization following the quasi optimality principle; the linear system in Eq.~\eqref{Eq:discrete_decon} was well-conditioned and hence directly solved without regularization.

\subsubsection{Extended dynamics}
\label{subsubsec:Extended_GLE}
Given the memory kernel ${K}(t)$ determined, the GLE in Eq. (\ref{Eq:gle2}) can be solved to predict the dynamics of the CG system. However, directly solving this equation requires to evaluate the time convolution of the memory kernel and velocity and to generate color noise for the random force, which needs to store the historical information and can be prohibitively expensive. Note that the solvent-mediated kinetics could result in a long-tailed memory kernel, making the computation even more expensive. To address the challenge of directly solving the GLE, $K(t)$ is first approximated by an asymptotic expansion as: 
\begin{equation}
K(t) \approx 
\sum\limits_{l=1}^{\mathcal{N}}\exp(-\frac{a_l}{2} t)[ b_l \cos(q_l t) + c_l \sin(q_l t)] \; ,
\label{equ:K_fit}
\end{equation}
where the parameters $\{a_l, b_l, c_l, q_l\}_{l=1,\dots,\mathcal{N}}$ can be determined via fitting \cite{li2017computing,Pan_CGMZ_2019}. Truncating the expansion with more terms (larger $\mathcal{N}$) leads to more accurate approximation of $K(t)$. Approximating the memory kernel by a finite set of exponentially damped oscillators as in Eq. \eqref{equ:K_fit} would allow to replace the GLE with a Markovian dynamics extended in higher dimensions. By doing so, the expensive cost of solving the GLE can be significantly reduced, as has been evidenced in literature \cite{ceriotti2010colored,li2017computing,Pan_CGMZ_2019}. To this end, Eq. \eqref{equ:K_fit} is rewritten in a matrix form as:
\begin{equation} \label{Eq:K_ApsEAsp}
    K(t) \approx 
    -\mathbf{A}_{ps} e^{-t\mathbf{A}_{ss}} \mathbf{A}_{sp} \; ,
\end{equation}
where $\mathbf{A}_{ps}=-\mathbf{A}_{sp}^T$. If we define the parameter matrix $\mathbf{A} = [0,~\mathbf{A}_{ps};~\mathbf{A}_{sp},~\mathbf{A}_{ss}]$, it can be assembled from the parameters in Eq. \eqref{equ:K_fit} by:
\begin{equation}
    \mathbf{A}_l  = 
  \left[
    \begin{array} {c:cr} 
        0	&	\sqrt[]{\frac{b_l}{2}- \frac{q_l c_l}{a_l} }	&	\sqrt[]{\frac{b_l}{2}+ \frac{q_l c_l}{a_l} }
        \\
        \\ \hdashline 
        \\
        -~\sqrt[]{\frac{b_l}{2}- \frac{q_l c_l}{a_l} }	&	a_l	&	\frac{1}{2} ~ \sqrt[]{4 q_l^2+a_l^2}	
        \\
       -~\sqrt[]{\frac{b_l}{2}+ \frac{q_l c_l}{a_l} } &	-~\frac{1}{2} ~ \sqrt[]{4 q_l^2+a_l^2}	& 0 
    \end{array}
  \right] \; .
  \label{Eq:assembleA}
\end{equation}
In Eq. \eqref{Eq:assembleA}, the top right block contributes to $\mathbf{A}_{ps}$; the bottom left contributes to $\mathbf{A}_{sp}$; and the block on the bottom right constitutes $\mathbf{A}_{ss}$, which is a block diagonal matrix consisting of $2\times2$ blocks. 

Given Eq. \eqref{Eq:K_ApsEAsp} and by introducing auxiliary variables $\mathbf{S}$, the extended Markovian  dynamics is given by:
	\begin{gather}\label{Eq:extend}
	\begin{pmatrix} \dot{\mathbf{P}}\\ \dot{\mathbf{S}} \end{pmatrix} = - \begin{pmatrix} 0&\mathbf{A}_{ps}\\ \mathbf{A}_{sp}&\mathbf{A}_{ss} \end{pmatrix}
	\begin{pmatrix} M^{-1}\mathbf{P}\\ \mathbf{S} \end{pmatrix} + 
	\begin{pmatrix} 0&0\\ 0&\mathbf{B}_{s} \end{pmatrix}
	\begin{pmatrix} 0\\  \boldsymbol{\xi} \end{pmatrix} \; .
	\end{gather}
Here, $\boldsymbol{\xi}$ is a vector of uncorrelated Gaussian random variables with $\langle \boldsymbol{\xi}(t)\rangle=\mathbf{0}$ and $\langle \xi_{I,\mu}(t)\xi_{J,\nu}(0) \rangle =\delta_{IJ}\delta_{\mu\nu}\delta(t)$, where $\xi_\nu$ and $\xi_\mu$ denote the different elements of $\boldsymbol{\xi}$. To satisfy the second fluctuation-dissipation theorem \cite{kubo1966r}, $\mathbf{B}_{s}\mathbf{B}_{s}^T  = k_BT(\mathbf{A}_{ss} + \mathbf{A}_{ss}^T)$. We can write the parameter matrix $\mathbf{B} = \text{diag}(0, \mathbf{B}_s)$. To ensure $\mathbf{A}$ and $\mathbf{B}$ are both real number matrices, the parameters in Eq. \eqref{equ:K_fit} need to satisfy: $a_l\geq 0$, $b_l \geq 0 $, and $|c_l | \leq \frac{a_l b_l}{2q_l}$. The extended dynamics in Eq. \eqref{Eq:extend} is equivalent to the GLE in Eq. \eqref{Eq:gle2} \cite{ceriotti2010colored} with the random force:
\begin{equation}
    \tilde {\mathbf{F}}(t) = -\int_{0}^{t} \mathbf{A}_{ps}e^{-(t-t')\mathbf{A}_{ss}} \mathbf{B}_{s} \boldsymbol{\xi}(t') dt' \;.
\end{equation}

The extended dynamics in Eq. \eqref{Eq:extend} circumvents the expensive time-convolution, samples only white noise, and hence can be solved more
efficiently than the GLE. In the present work, the implicit velocity-Verlet temporal integrator \cite{brunger1984stochastic} was used to numerically solve Eq. \eqref{Eq:extend} in the CG simulations.

\subsection{Transfer learning of memory kernel}\label{subsec:transfer_learning}
In this section, we explain the methodology proposed for transfer learning of memory kernel, which enables the CG modeling transferable across a range of parameters to reproduce the dynamic properties of the underlying atomistic systems. The proposed methodology draws on the GPR and can be significantly accelerated by the model order reduction and active learning techniques.

\subsubsection{Transfer learning based on GPR}\label{subsubsec:GPR}

The memory kernel is denoted as $K({t};\boldsymbol \mu)$, where the vector $\boldsymbol \mu$ represents the parameters of interest such as temperature, concentration, and solvent viscosity; and $({t},\boldsymbol \mu) \in \mathcal{T} \times \mathcal{P}$ with $\mathcal{T}  = [0,t_f] $ and $ \mathcal{P} \in \mathbb{R}^d$ representing the time domain and parameter space, respectively, where $d$ is the dimension of the parameter space. Our goal is to predict the memory kernel $K({t};\boldsymbol \mu^*)$ at any given parameter instance $\boldsymbol \mu^* \in \mathcal{P}$ by using the data of $K({t};\boldsymbol \mu)$ at $N_p$ training parameter instances. To achieve this goal, we define $\mathbf{x} = [t,\boldsymbol \mu] \in \mathbb{R}^{d+1}$ and $y = K(\mathbf{x}) \in \mathbb{R}$ as the input and output of GPR, respectively. Recalling that the data of memory kernel are obtained at $N_t$ discrete times (\S\ref{subsubsec:memory}), we totally have $N_\text{train} = N_t \times N_p$ training data for GPR. The dependence of the outputs (memory kernels) $\mathbf{Y} = [ y_1, y_2,\dots , y_{N_\text{train}} ]^T \in \mathbb{R}^{N_\text{train}}$ on the inputs (time and parameters) $\mathbf{X} = [ \mathbf{x}_1, \mathbf{x}_2,\dots , \mathbf{x}_{N_\text{train}} ]^T \in \mathbb{R}^{N_\text{train} \times (d+1)}$ is modeled as a Gaussian process: $\mathbf{Y} (\mathbf{X}) \sim  \mathcal{GP}(\mathbf{u} (\mathbf{X} ),\boldsymbol \Sigma (\mathbf{X} ,\mathbf{X}))$. Here, $\mathbf{u}(\mathbf{X}) \in \mathbb{R}^{N_\text{train}}  $ is the mean function, and $\boldsymbol \Sigma (\mathbf{X} ,\mathbf{X}) \in \mathbb{R}^{N_\text{train} \times N_\text{train}} $ is the covariance matrix. In this work, the covariance function is assumed a squared exponential form, i.e.,
\begin{equation}
    \Sigma_{ij}( \mathbf{X},\mathbf{X}; \boldsymbol{\theta})=\theta_f^2 \exp \left [-\frac{1}{2} \sum\limits_{k=1}^{d+1}  \theta_{l_k}^2({x}_{i,k} -  {x}_{j,k})^2 \right ],
\end{equation}
where $ {x}_{i,k}$ and $ {x}_{j,k}$ refer to the $k$-th element of $ \mathbf{x}_i$ and $ \mathbf{x}_j$, respectively; $\boldsymbol{\theta} = (\theta_f, \theta_{l_1}, \theta_{l_2}, \dots, \theta_{l_{d+1}})$ denotes the hyper-parameters. Using the GPR model inferred from the training data, we can predict $K({t};\boldsymbol \mu^*)$ at any given $\boldsymbol \mu^*$. For that, the inputs are $ \mathbf{X}^* = [ \mathbf{x}^*_1, \mathbf{x}^*_2,\dots , \mathbf{x}^*_{N_t} ] $ with $\mathbf{x}^*_i = [t_i, \boldsymbol \mu^*] $; the outputs are $\mathbf{Y}^*=[K(t_1;\boldsymbol \mu^*),K(t_2;\boldsymbol \mu^*),\dots,K(t_{N_t};\boldsymbol \mu^*)]^T$ and 
satisfy:
\begin{equation}
    \mathbf{Y}^*|\mathbf{Y} \sim \mathcal{GP}(\hat{\mathbf{u}},\hat{\boldsymbol \Sigma }) \;,
\end{equation}
where
\begin{equation}\label{Eq:posterior_muC}
\begin{split}
    \hat{\mathbf{u}}&=\boldsymbol \Sigma (\mathbf{X}^*,\mathbf{X} )\left [\boldsymbol \Sigma (\mathbf{X} ,\mathbf{X}) +\sigma^2\mathbf{I}\right]^{-1}\left [\mathbf{Y} - \mathbf{u} (\mathbf{X} ) \right ] + \mathbf{u} (\mathbf{X}^* ),\\
    \hat{\boldsymbol \Sigma }&=\boldsymbol \Sigma (\mathbf{X}^*,\mathbf{X}^* )-\boldsymbol \Sigma (\mathbf{X}^*,\mathbf{X} )\left [\boldsymbol \Sigma (\mathbf{X} ,\mathbf{X} )+\sigma^2\mathbf{I}\right]^{-1}\boldsymbol \Sigma (\mathbf{X}^*,\mathbf{X} )^T \;,
\end{split}
\end{equation}
and $\sigma^2$ is the variance of identically independent Gaussian noise (with zero mean) assumed in the GPR model. In Eq.~\eqref{Eq:posterior_muC}, $\hat{\mathbf{u}}$ is the mean value of the prediction at $\boldsymbol \mu^*$; the diagonal elements of $\hat{\boldsymbol \Sigma }$ are the analytical uncertainty bounds, denoted as $\hat \sigma$. To determine $\sigma^2$ and the hyper-parameters $\boldsymbol{\theta}$, we minimize the negative log marginal likelihood \cite{GPML_2006}:
\begin{equation}\label{Eq:marginal_likelihood}
    -\log p(\mathbf{Y} |\boldsymbol{\theta}, \sigma^2)=\frac{1}{2}\mathbf{Y} ^T\mathbf{\mathcal{C}}^{-1} \mathbf{Y} +\frac{1}{2} \log |\mathbf{\mathcal{C}} | + \frac{N_\text{train}}{2} \log (2\pi) \; ,
\end{equation}
using the Quasi-Newton optimizer L-BFGS \cite{BFGS_1989Liu}, where $\mathbf{\mathcal{C}} = \boldsymbol \Sigma (\mathbf{X} ,\mathbf{X})+\sigma^2\mathbf{I}$. Once $\sigma^2$ and the hyper-parameters are determined, the GPR model is complete and can be used to predict the memory kernel at any parameter instance of interest along with the uncertainty quantified for the prediction.

\subsubsection{Transfer learning accelerated by model order reduction}\label{subsubsec:POD}
The training cost of GPR exhibits a cubic scaling with respect to the number of training data, i.e. $\mathcal{O}(N_\text{train}^3)$. In practice, $N_\text{train} = N_p \times N_t$ can be a large number due to the large $N_t$. As we have discussed above, the dynamics of a polymer in solution can display a long-tailed memory kernel. To fully capture the memory kernel's variations in all time scales, the data must be collected at many discrete times, resulting in large $N_t$. In the numerical examples considered in this work, $N_t$ is larger than $N_p$ (number of training parameter instances) by 2 or 3 orders. Hence, if the memory kernel is directly modeled by GPR as discussed in \S\ref{subsubsec:GPR}, the training cost can be expensive due to a large $N_\text{train} $. Thus, we further propose a strategy to greatly accelerate the transfer learning by combining the GPR with model order reduction. In particular, the memory kernel is first decomposed using POD into temporal and parameter modes, from which only the dominant modes are retained. GPR is then used to model the dominant parameter modes only.

Same as in \S\ref{subsubsec:GPR}, we need snapshot data of $K({t};\boldsymbol \mu_i)$ at $\boldsymbol \mu_i$ with $i=1, 2, \dots, N_p$. A reduced order model (ROM) is established for the memory kernel by decomposing the snapshot data of $K({t};\boldsymbol \mu_i)$ as:
\begin{equation}\label{Eq:snap_decompose}
K({{ t}} ; \boldsymbol \mu_i) = \bar K ({{ t}}) + \hat K({{ t}} ; \boldsymbol \mu_i) = \bar K ({{t}}) + \sum\limits_{k = 1}^{N_p} \alpha_k(\boldsymbol \mu)\phi_k(t) \;,
\end{equation}
where $\bar K(t) =  \frac{1}{N_p} \sum\limits_{i = 1}^{N_p} K({{ t}} ; \boldsymbol \mu_i)$ denotes the mean of all snapshots; the fluctuating part $\hat K$  is decomposed into the temporal bases $\phi_k({t})$ and the parameter modes $\alpha_k(\boldsymbol \mu)$. Note that in the classical model order reduction for dynamical systems, the POD basis functions are spatial bases \cite{Ma2021PODGPR}. Here, we adapt the technique to our needs and replace the spatial bases with temporal bases. According to POD, the basis functions can be obtained by eigendecomposition of the correlation matrix $\mathbf{G} \in \mathbb{R}^{N_p\times N_p}$ of the fluctuating parts:
\begin{equation}\label{Eq:A_correlation}
{G_{ij}} = \int_0^{t_f}  {{{\hat K}}({{t}}; \boldsymbol \mu_i){{\hat K}}({{t}} ;\boldsymbol \mu_j)dt} \;,
\end{equation}
where $i$ and $j$ refer to the $i$-th and $j$-th snapshots, respectively. 
Then, the temporal bases $\phi_k(t)$ (i.e., POD basis functions) are given as:
\begin{equation}\label{Eq:phi_k}
    {\phi _k}(t) = \frac{1}{{\sqrt {{\lambda _k}} }}\sum\limits_{i = 1}^{N_p} {w_i^k\hat K({{ t}} ; \boldsymbol \mu_i) } \;,
\end{equation}
where $\{\lambda_1, \lambda_2, ..., \lambda_M\}$ are the eigenvalues of  $\mathbf{G}$ in descending order; $w_i^k$ is the $i$-th component of $\mathbf{w}^k$, the eigenvector corresponding to the eigenvalue $\lambda_k$. 
The idea of POD is that the energy contributed by  each basis can be reflected by its corresponding eigenvalue. Thus, if the first $R \ll N_p$ eigenvalues are significantly larger than the remaining, we only need to retain the first $R$ modes since they dominate the energy, i.e., 
\begin{equation}\label{equ:K_POD_trunc}
    \hat K(t;\boldsymbol \mu_i) = \sum\limits_{k = 1}^{N_p} \alpha_k(\boldsymbol \mu)\phi_k(t)
    \approx 
    \sum\limits_{k = 1}^R \alpha_k(\boldsymbol \mu_i) \phi_k({{t}}) = \sum\limits_{k = 1}^R \sqrt {{\lambda _k}}{w_k^i} \phi_k({{t}})  \; ,
\end{equation}
where $\alpha_k(\boldsymbol \mu_i) = \sqrt {{\lambda _k}}{w_k^i} $. The approximation (truncation) error of Eq.~\eqref{equ:K_POD_trunc} is $\epsilon^\text{POD} = \sqrt{\sum\limits_{k=R+1}^{N_p} \lambda_k}$, which can be reduced by increasing $R$, i.e., by keeping more POD modes. 
Based on it, we define the relative error of POD as:
\begin{equation}\label{Eq:RRMS}
    \epsilon_r^{\text{POD}} = \sqrt{ \frac{\sum\limits_{k=R+1}^{N_p} \lambda _k}{\sum\limits_{k=1}^{N_p} \lambda_k} } \;.
\end{equation}
For a target tolerance $\zeta_\text{POD}$, by requiring $\epsilon_r^{\text{POD}} \leq \zeta_\text{POD} $, we can determine how many dominant modes to retain in the ROM, i.e., the value of $R$. 

Given Eq.~\eqref{equ:K_POD_trunc}, we can establish the ROM for the memory kernel as:
\begin{equation}\label{Eq:ROM_POD}
K(t;\boldsymbol \mu^*) \approx K^\text{ROM}(t;\boldsymbol \mu^*)= \bar K (t) + \sum\limits_{k = 1}^R \alpha_k(\boldsymbol \mu^*) \phi_k({{t}}) \; ,
\end{equation}
by which the memory kernel can be effectively predicted at a given parameter instance $\boldsymbol \mu^*$ by the reduced temporal bases $\phi_k({t})$ and parameter modes $\alpha_k(\boldsymbol \mu^*)$. Thus, we only need to train and predict $\alpha_k(\boldsymbol \mu^*)$ instead of the entire memory kernel function $K(t;\boldsymbol \mu^*)$, resulting in greatly reduced training and prediction costs. Following the idea of data-driven nonintrusive reduced order modeling \cite{Ma2021PODGPR}, we still use GPR to model each parameter mode $\alpha_k(\boldsymbol \mu)$, i.e., 
$\alpha_k(\boldsymbol \mu) = y (\mathbf{x}) \sim  \mathcal{GP}( u_k(\mathbf{x} ), \Sigma_k (\mathbf{x} ,\mathbf{x}^*))$, where the input and output of GPR become $\mathbf{x} = \boldsymbol \mu \in \mathbb{R}^{d}$ and $y = \alpha_k(\boldsymbol \mu)$, respectively. Now the number of training data required in GPR is $N_\text{train} = N_p$, which is much smaller than $N_\text{train} = N_p \times N_t$ in \S\ref{subsubsec:GPR}. Once the GPR model of each parameter mode is constructed, we can predict  $\alpha_k(\boldsymbol \mu^*)$ and in turn the memory kernel by Eq.~\eqref{Eq:ROM_POD} at any given parameter instance $\boldsymbol \mu^*$. In addition, the uncertainty for the predicted $\alpha_k(\boldsymbol \mu^*)$ can be quantified by the standard deviation $\hat \sigma_k(\boldsymbol \mu^*)$ in the GPR model. 


\subsubsection{Active learning with minimum training data}\label{subsubsec:active_learning}

To ensure the accuracy of the inferred GPR models, sufficient training data are required. However, attaining the data of memory kernels needs nontrivial efforts. Thus, minimizing the number of data required for training the GPR models becomes necessary. To this end, we propose an active learning strategy to adaptively sample the training data and to ensure maximum information gain from the data sampled. Its key idea is to take the advantage of the fact that the GPR can predict not only the mean $\hat u $ but also the standard deviation $\hat \sigma$ to quantify the uncertainty. The knowledge of uncertainty can guide the sampling of the next data. In particular, the next parameter instance $\boldsymbol{\mu}_\text{next}$ to be sampled is where the uncertainty level reaches maximum over the sampling space. In this work, active learning is only integrated with the transfer learning accelerated by model order reduction (\S\ref{subsubsec:POD}), but not with the transfer learning solely based on GPR (\S\ref{subsubsec:GPR}). This is because in \S\ref{subsubsec:GPR} the computational cost for updating the GPR model after each new sampling is significant, which can overshadow the benefit of active learning. The proposed procedure is described as below. 

Let $\mathcal{P}_\text{tr} \in \mathbb{R}^d$ denote the training sampling space and note $\mathcal{P}_\text{tr} \subseteq \mathcal{P}$, where $\mathcal{P}$, as mentioned before, represents the entire parameter space considered including the region beyond the range of training sampling space. The active learning starts with a small set of sampling points, denoted as $\mathcal{P}^\text{active}_{\text{tr}}$, which can be chosen randomly or uniformly over the training sampling space. The data of memory kernels $K(t;\boldsymbol \mu)$ for all $\boldsymbol \mu \in \mathcal{P}^\text{active}_{\text{tr}}$ are generated following \S\ref{subsubsec:memory}. After that, the following steps are taken successively: 1) build the ROM for the memory kernel as in Eq.~\eqref{Eq:ROM_POD} and construct a GPR model for each $\alpha_k(\boldsymbol \mu)$, $k=1, 2, \dots, R$; 2) quantify the uncertainty $\bar \sigma (\boldsymbol \mu)$ of the GPR models for each $\boldsymbol \mu \in \mathcal{P}_\text{tr}$ as: 
\begin{equation}\label{Eq:uncertainty}
    \bar \sigma(\boldsymbol \mu ) =  \frac{\sum\limits_{k=1}^R \hat \sigma_k(\boldsymbol \mu)}{\sum\limits_{k=1}^R \|\hat{\mathbf{u}}_k\|_\infty}  \;,
\end{equation}
with $\hat \sigma_k(\boldsymbol \mu )$ the standard deviation of the GPR model for $a_k(\boldsymbol \mu )$ and $ \| \hat{\mathbf{u}}_k\|_\infty$ the $L_\infty$ norm of $\hat{\mathbf{u}}_k$, a vector consisting of $\hat u_k$ for each $\boldsymbol \mu \in \mathcal{P}_{\text{tr}}$; 3) determine the next sampling parameter instance: $\boldsymbol \mu_\text{next} = \argmax\limits_{\boldsymbol \mu \in \mathcal{P}_{\text{tr}} }\bar \sigma (\boldsymbol \mu)$; and 4) generate the data of the memory kernel $K(t;\boldsymbol{\mu}_\text{next})$ at $\boldsymbol\mu_\text{next}$ and augment the training data set with the new data. These four steps are repeated until the maximum uncertainty $\bar \sigma(\boldsymbol \mu_\text{next})$ is less than the preset tolerance $\zeta_\text{AL}$. By such, the active learning is able to start with a small set of training data and then to adaptively add more data as necessary at locations that can maximize information gain. The entire procedure is also outlined in Algorithm \ref{Alg:active_learning}. 
\begin{algorithm}
\caption{Active learning}\label{Alg:active_learning}
\begin{algorithmic}
\State Generate data of $K(t;\boldsymbol \mu)$ for all $\boldsymbol \mu \in \mathcal{P}^\text{active}_{\text{tr}}$
\State Set $\zeta_\text{AL}$, $\zeta_\text{POD}$, and $\bar \sigma(\boldsymbol \mu_\text{next}) = \infty$
\While{$\bar \sigma(\boldsymbol \mu_\text{next}) > \zeta_\text{AL}$}
\State Build the ROM as in Eq.~\eqref{Eq:ROM_POD} via POD with the tolerance $\zeta_\text{POD}$
\State Construct a GPR model for each $\alpha_k(\boldsymbol \mu)$, $k = 1,\dots,R$
\State Quantify the uncertainty $\bar \sigma (\boldsymbol \mu)$ of the GPR models for each  $\boldsymbol \mu \in \mathcal{P}^\text{tr}$
\State Determine $ \boldsymbol \mu_\text{next} = \argmax\limits_{\boldsymbol \mu \in \mathcal{P}_{\text{tr}} } \bar \sigma (\boldsymbol \mu)$
\State Generate the data of $K(t;\boldsymbol{\mu}_\text{next})$
\State Augment the training parameter instances with $ \mathcal{P}^\text{active}_{\text{tr}} = \mathcal{P}^\text{active}_{\text{tr}} \cup  \{\boldsymbol \mu_\text{next}\} $
\EndWhile
\State Output the ROM (as in Eq. \eqref{Eq:ROM_POD}) with the temporal bases $\phi_k(t)$ and the GPR model for each $\alpha_k(\boldsymbol \mu)$, $k = 1,\dots,R$
\end{algorithmic}
\end{algorithm}
\section{Results}\label{sec:results}
In this section, we assessed the accuracy and efficiency of our proposed transfer learning methods in two example systems: star polymer and peptoid polymer. The dynamics of star polymer or peptoid polymer in solution can be affected by the parameters such as temperature, concentration, and solvent property. We aimed to establish transferable memory kernels so that the CG models can faithfully reproduce the dynamic properties of the reference atomistic systems across different values of the parameters. The dynamic properties include the VACF, diffusion coefficient, and mean square displacement (MSD) as functions of time. For convenience, we hereinafter denote the transfer learning method solely based on GPR (\S\ref{subsubsec:GPR}) as GPR transfer learning and the transfer learning method via both ROM and GPR (\S\ref{subsubsec:POD}) as ROM-GPR transfer learning.  

In CG modeling, each star polymer or peptoid polymer was coarse-grained as a single CG particle; the solvent (or water) DOFs were all eliminated. The memory kernels computed from atomistic data were obtained following \S\ref{subsubsec:memory}. To construct a CG model, the memory kernel predicted by transfer learning was approximated by Eq.~\eqref{equ:K_fit}. The dynamics of the CG system was simulated by solving Eq. \eqref{Eq:extend} numerically using the implicit velocity-Verlet temporal integrator, for which in-house computer codes were developed under the framework of LAMMPS \cite{plimpton1995fast}.

\subsection{Star-polymer solution}
\begin{figure}[H]
    \centering
    \includegraphics[width=3.5cm]{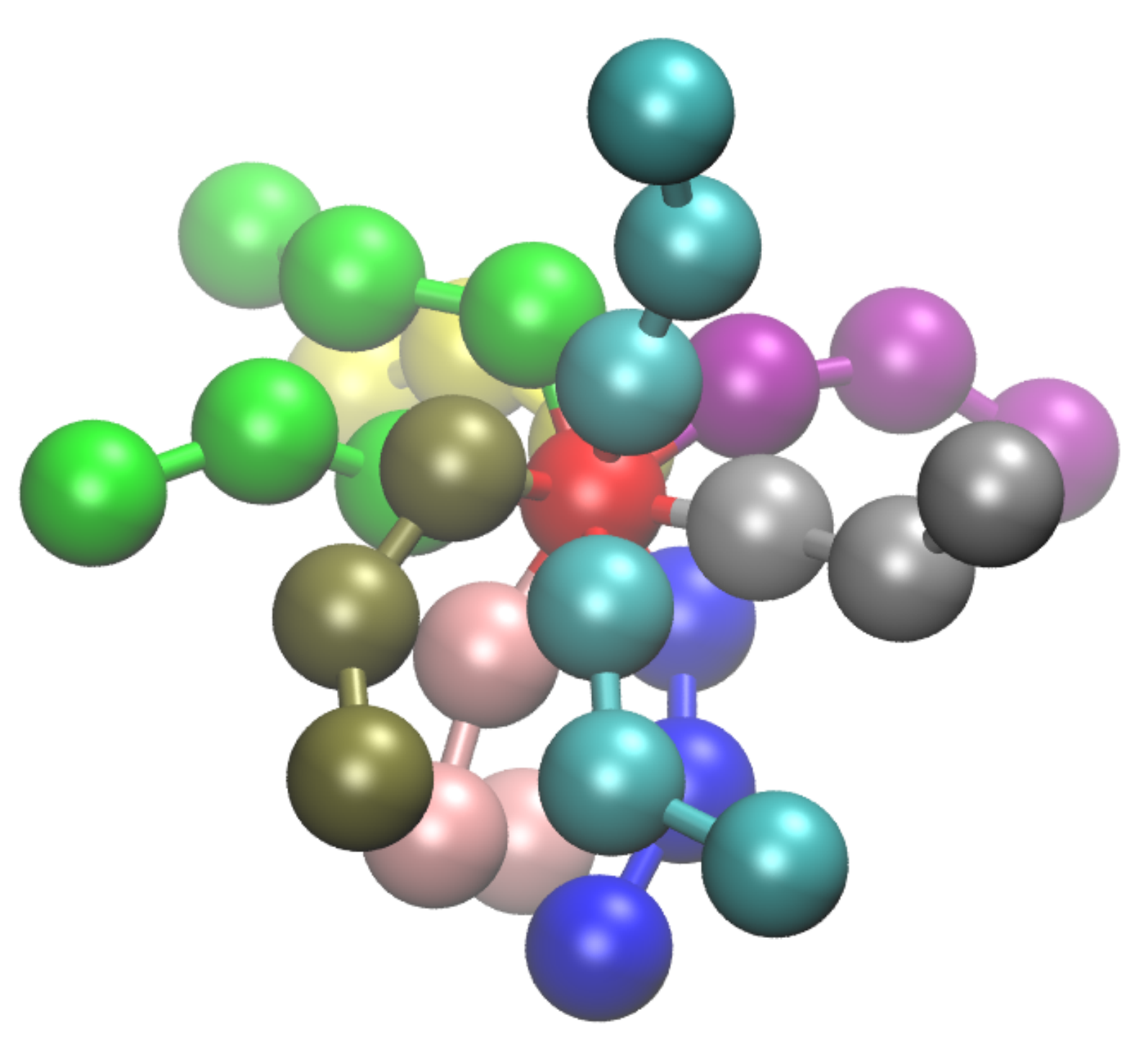}
    \caption{Atomistic model of a star polymer consisting of 31 LJ beads: one core and 10 arms with 3 beads per arm.}
    \label{fig:N31_allatom}
\end{figure}

We first considered solutions of stat polymers. In the atomistic representation, each star polymer consists of a core Lennard-Jones (LJ) bead and 10 identical arms with 3 LJ beads per arm, as illustrated in Fig.~\ref{fig:N31_allatom}. 
The core LJ bead and the LJ beads in each arm are identical and connected by the finitely extensible non-linear elastic (FENE) bonds. The solvent is also modeled by LJ beads. The dynamics of the atomistic system is governed by the Hamiltonian:
\begin{equation}\label{equ:Hamiltonian}
H = \sum\limits_{i=1}^{n}\frac{\textbf{p}_i^2}{2m_i}+\sum\limits_{i\neq j} E(r_{i j}) \; ,
\end{equation}
where $n$ is the total number of LJ beads in the atomistic system; $\textbf{p}_i$ and $m_i$ are the momentum and mass of the $i$-th LJ bead, respectively; $r_{i j}=|\textbf{r}_{i j}|=| \textbf{r}_i - \textbf{r}_j|$ is the distance between two LJ beads; and $E$ denotes the total potential energy contributed by the interatomic and bonded potentials. 

The interatomic LJ potential between the $i$-th and $j$-th LJ beads adopts the purely repulsive Weeks-Chandler-Andersen (WCA) potential and is given by: 
\begin{equation}\label{WCA}
 E_\text{WCA}(r_{ij}) =
    \begin{cases}
      4\epsilon_{ij}[(\frac{\sigma_{ij}}{r_{ij}})^{12}-(\frac{\sigma_{ij}}{r_{ij}})^6+\frac{1}{4}] & r_{ij} \leq r_\text{cut}\\
      \infty & r_{ij} > r_\text{cut}\\
    \end{cases}  \; , 
\end{equation}
where $r_\text{cut} = 2^{1/6}\sigma_{ij}$ is the cutoff distance. We denote the WCA potential's parameters for the star polymers as $\epsilon_\text{star}$ and $\sigma_\text{star}$ and for the solvent as $\epsilon_\text{sol}$ and $\sigma_\text{sol}$. In all MD simulations, while the values of $\epsilon_\text{star}$, $\sigma_\text{star}$, and $\epsilon_\text{sol}$ were fixed as $\epsilon_\text{star}=1$, $\sigma_\text{star}=1$, and $\epsilon_\text{sol}=1$, the value of $\sigma_\text{sol}$ varied for different solution systems between the range $\sigma_\text{sol}\in [1.0, 2.2]$. For the interaction between a star-polymer LJ bead and a solvent LJ bead, the WCA potential's parameters were determined by the geometric mix rule.

The FENE potential for the bonded interaction between connected LJ beads in each star polymer is:
\begin{equation}\label{FENE}
 E_\text{FENE}(r_{ij}) =
    \begin{cases}
      -\frac{1}{2}k_{b}r_0^2\ln[1-(\frac{r_{ij}}{r_0})^2] & r_{ij} \leq r_0\\
      \infty & r_{ij} > r_0\\
    \end{cases} \; ,   
\end{equation}
where $k_{b}=30$ is the spring constant, and $r_0=1.5$ is the maximum length of the FENE spring. 

Here, the reduced LJ units were employed; the mass of all LJ beads was chosen to be unity. Each star-polymer solution consists of 620,000 LJ beads. 
The MD simulations were performed using LAMMPS \cite{plimpton1995fast}. In all simulations, a periodic cubic box of length 115.7295 was used, which is large enough to neglect the effect of periodic box on the VACF. The Nose-Hoover thermostat under the canonical ensemble (NVT) was employed with the time step $\Delta t=0.001$. In each MD simulation, the data after reaching thermal equilibrium were collected for computing the ensemble-averaged quantities of interest. 

The concentration of star polymers in each solution was different and varied with $\gamma\in [0.2, 0.8]$, where $\gamma$ is defined as the number of star-polymer LJ beads divided by the total number of all LJ beads. In addition, the temperature varied with $T \in [0.5, 2.0]$ for each system. Thus, there are three parameters that differentiate the star-polymer solution systems considered herein, i.e., the concentration of star polymers $\gamma$, the temperature $T$, and the solvent viscosity characterized by $\sigma_\text{sol}$. 
We applied the proposed transfer learning methods to enable the memory kernel transferable across these parameters. The memory kernel was determined up to $t_f = 20$ until the VACF ($C(t_f)$) decayed to $ |C(t_f) |/|C(0) |\leq 10^{-2}$, which is long enough to capture the dynamics in all time scales. 

To compare the two methods (GPR transfer learning and ROM-GPR transfer learning) and also to demonstrate how the proposed active learning technique can be implemented, we first limited the transfer learning to one parameter. Thereafter, we tackled the more challenging case over all three parameters using the ROM-GPR transfer learning method along with active learning.

\subsubsection{Transferable in one parameter}
\label{subsubsec:N31-1d}
The temperature and solvent viscosity were fixed at $T = 1.0$ and $\sigma_\text{sol} = 1.0$, respectively. The concentration of star polymers $\gamma$ was the only changing parameter, i.e., $ \boldsymbol{\mu} = \gamma \in [0.2,0.8]$. Hence, the parameter space is one dimensional: $\mathcal{P}= [0.2,0.8] $. Here, the parameter instances used for training were from $\mathcal{P}_\text{tr} = [0.2,0.6]$. The proposed active learning technique allowed to adaptively sample the training data and to minimize the data needed for training the GPR models given the target tolerance of accuracy. 

\begin{figure}[t]
    \centering
    \includegraphics[width=8.8cm]{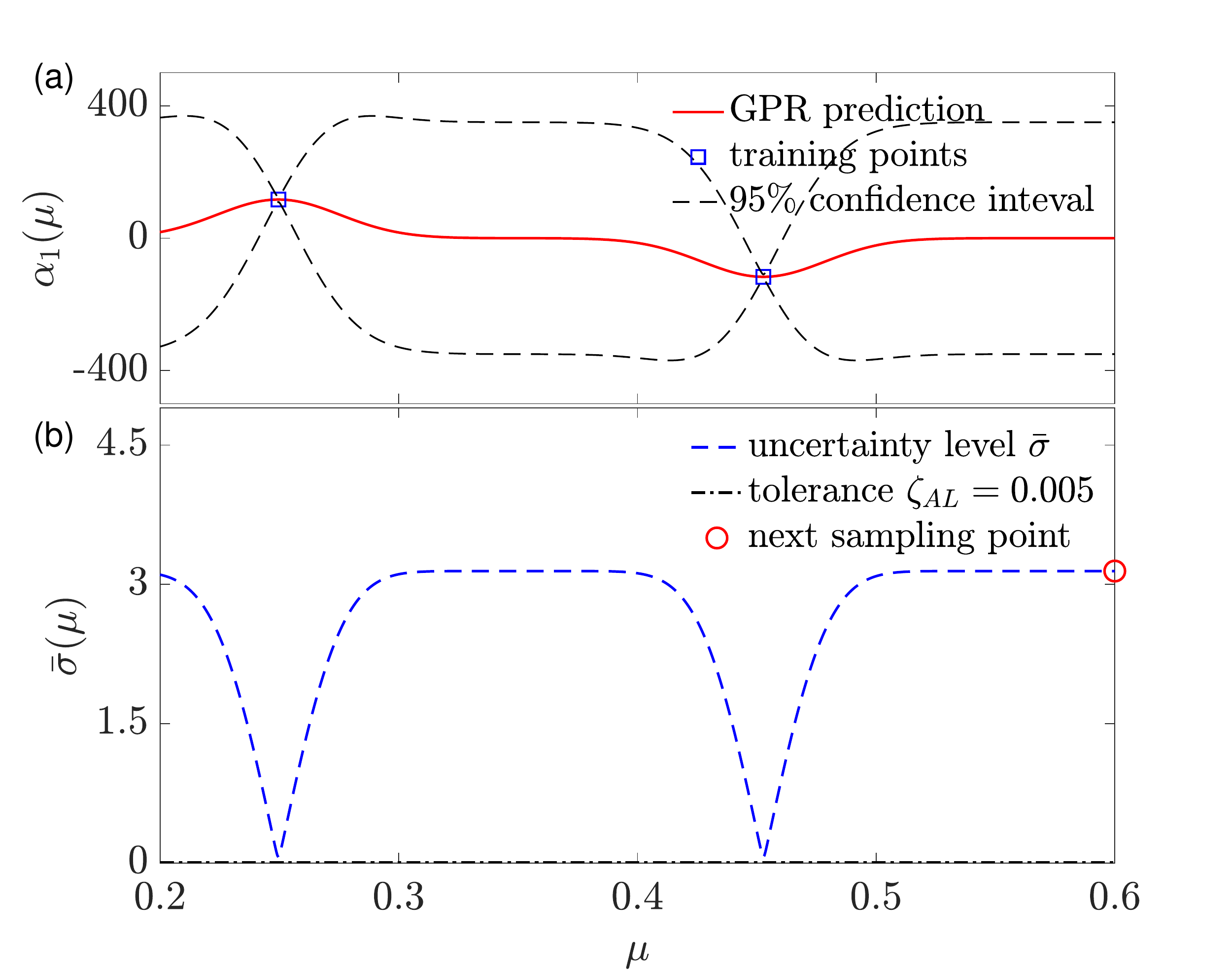}
    \caption{(a) GPR model for the POD parameter mode $\alpha_1(\mu)$ inferred from 2 training data. (b) Uncertainty quantified by $\bar \sigma(\mu)$ of the inferred GPR model and the next sampling point determined by active learning.}  
    \label{fig:First_GPR}
\end{figure}

\begin{figure}[htbp]
    \centering
    \includegraphics[width=8.8cm]{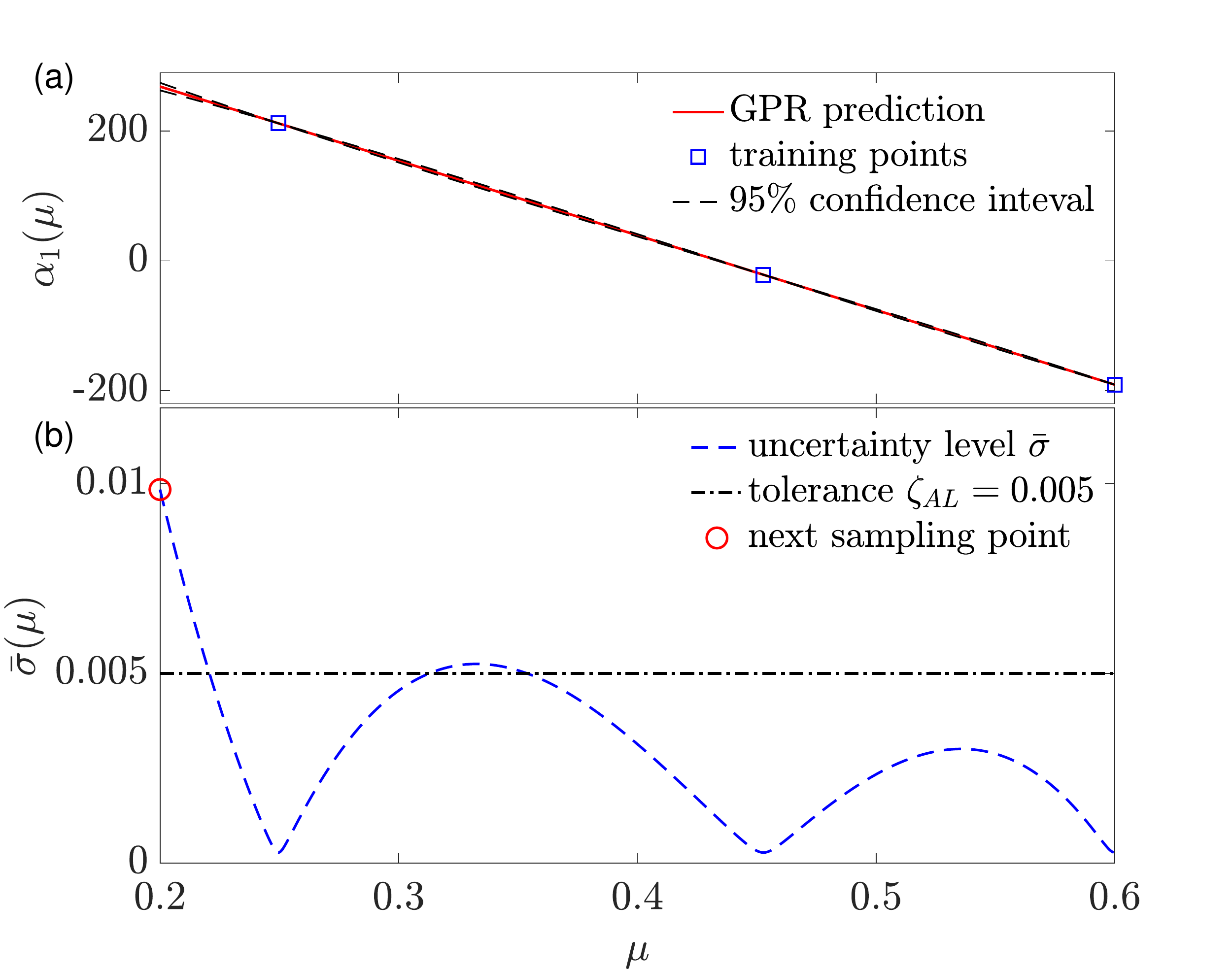}
    \caption{(a) GPR model for the POD parameter mode $\alpha_1(\mu)$ inferred from 3 training data. (b) Uncertainty quantified by $\bar \sigma(\mu)$ of the inferred GPR model and the next sampling point determined by active learning.}  
    \label{fig:Second_GPR}
\end{figure}

\begin{figure}[htbp]
    \centering
    \includegraphics[width=8.8cm]{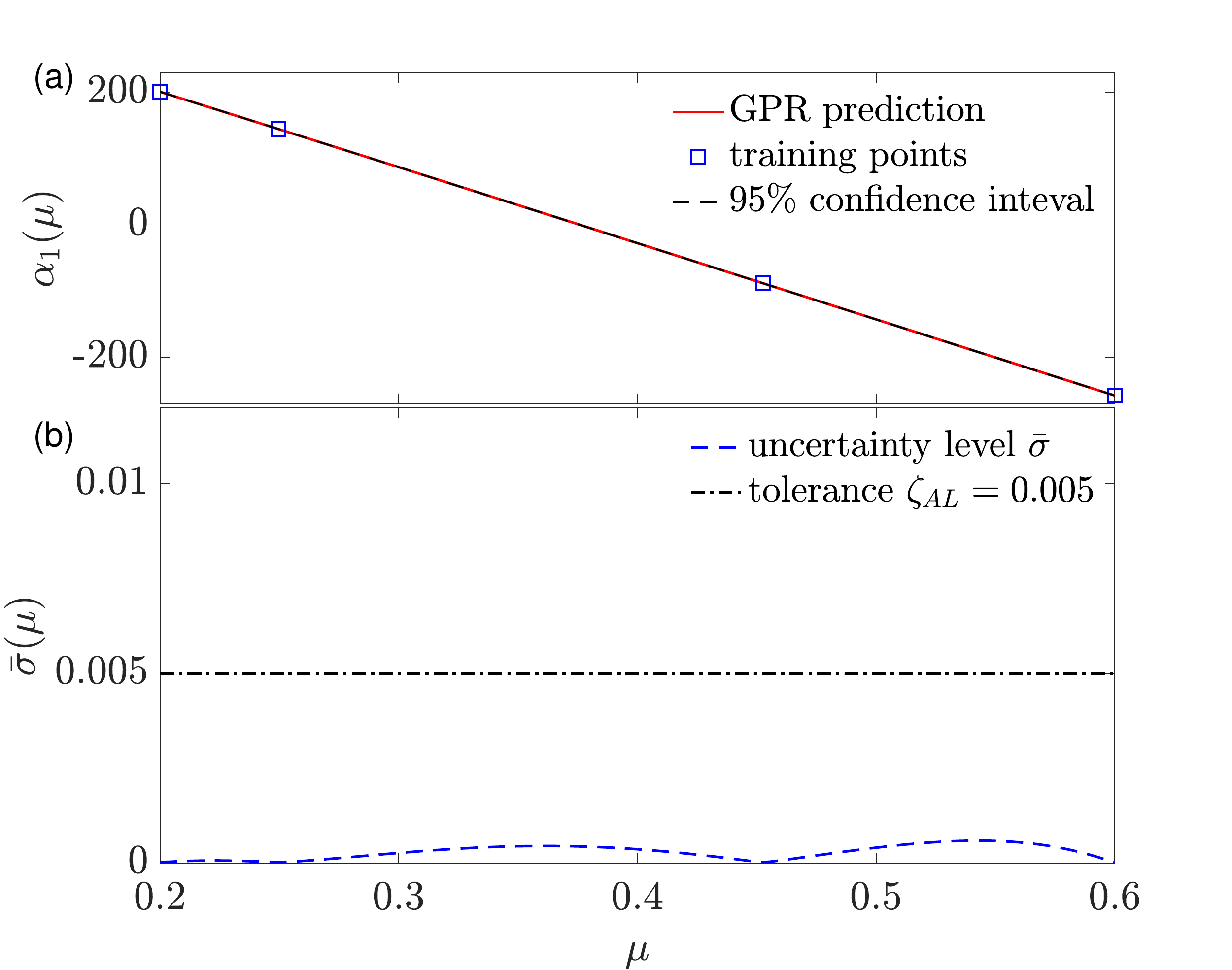}
    \caption{(a) GPR model for the POD parameter mode $\alpha_1(\mu)$ inferred from 4 training data. (b) Uncertainty quantified by $\bar \sigma(\mu)$ of the inferred GPR model.}  
    \label{fig:Third_GPR}
\end{figure}

The tolerances for POD and active learning were set as $\zeta_\text{POD} = 0.1$ and $\zeta_\text{AL} = 0.005$, respectively. The active learning process was initiated with two initial parameter instances randomly sampled as $\mathcal{P}^\text{active}_\text{tr} = \{ 0.2496, 0.4528 \}$. At these two parameter instances, we computed the memory kernels $K(t;\gamma=0.2496)$ and $K(t;\gamma=0.4528)$ from the MD simulation data as described in \S\ref{subsubsec:memory}. Through POD we constructed the ROM as in Eq.~\eqref{Eq:ROM_POD}, where $R=1$ and only the first POD basis needs to be retained for the target $\zeta_\text{POD} = 0.1$. The GPR model for the parameter mode $\alpha_1(\mu)$ inferred from the two training data, along with its uncertainty as defined in Eq.~\eqref{Eq:uncertainty}, are shown in Fig.~\ref{fig:First_GPR}. With only two training data, the inferred GPR model exhibits large uncertainty, which indicates more training data are needed. Since the uncertainty quantified by $\bar \sigma(\mu)$ reaches its maximum at $\mu = 0.6$, it was chosen as the next sampling point $\mu_\text{next} = 0.6$. The memory kernel $K(t;\gamma=0.6)$ was computed from the MD simulation data and added to the training data.  
Now the training sampling set became $ \mathcal{P}^\text{active}_\text{tr} = \{ 0.2486, 0.4528, 0.6 \}$. With the three training data, the GPR model inferred for $\alpha_1(\mu)$ along with its uncertainty are depicted in Fig.~\ref{fig:Second_GPR}. Compared with Fig.~\ref{fig:First_GPR}, the uncertainty of the GPR model was greatly reduced by adding one more training data. However, the maximum uncertainty is still larger than the preset tolerance $\zeta_\text{AL} = 0.005$, and hence, the next sampling point was chosen at where the maximum uncertainty occurs, i.e., $\mu_\text{next} = 0.2$. 
With that, the training sampling set became $ \mathcal{P}^\text{active}_\text{tr} = \{ 0.2486, 0.4528, 0.6, 0.2 \} $. The GPR model for $\alpha_1(\mu)$ and its uncertainty inferred from the four training data are illustrated in Fig.~\ref{fig:Third_GPR}. By including one more training data, the GPR model's uncertainty was further reduced, and the maximum uncertainty was less than the preset tolerance $\zeta_\text{AL} = 0.005$. Thus, the sampling process via active learning could be terminated.
The GPR model for $\alpha_1(\mu)$ in Fig.~\ref{fig:Third_GPR} was then substituted into the ROM in Eq.~\eqref{Eq:ROM_POD} with $R=1$ to predict the memory kernel at any given parameter $\mu^* \in \mathcal{P}$.

\begin{figure}[htbp]
    \centering
    \includegraphics[width=8.8cm]{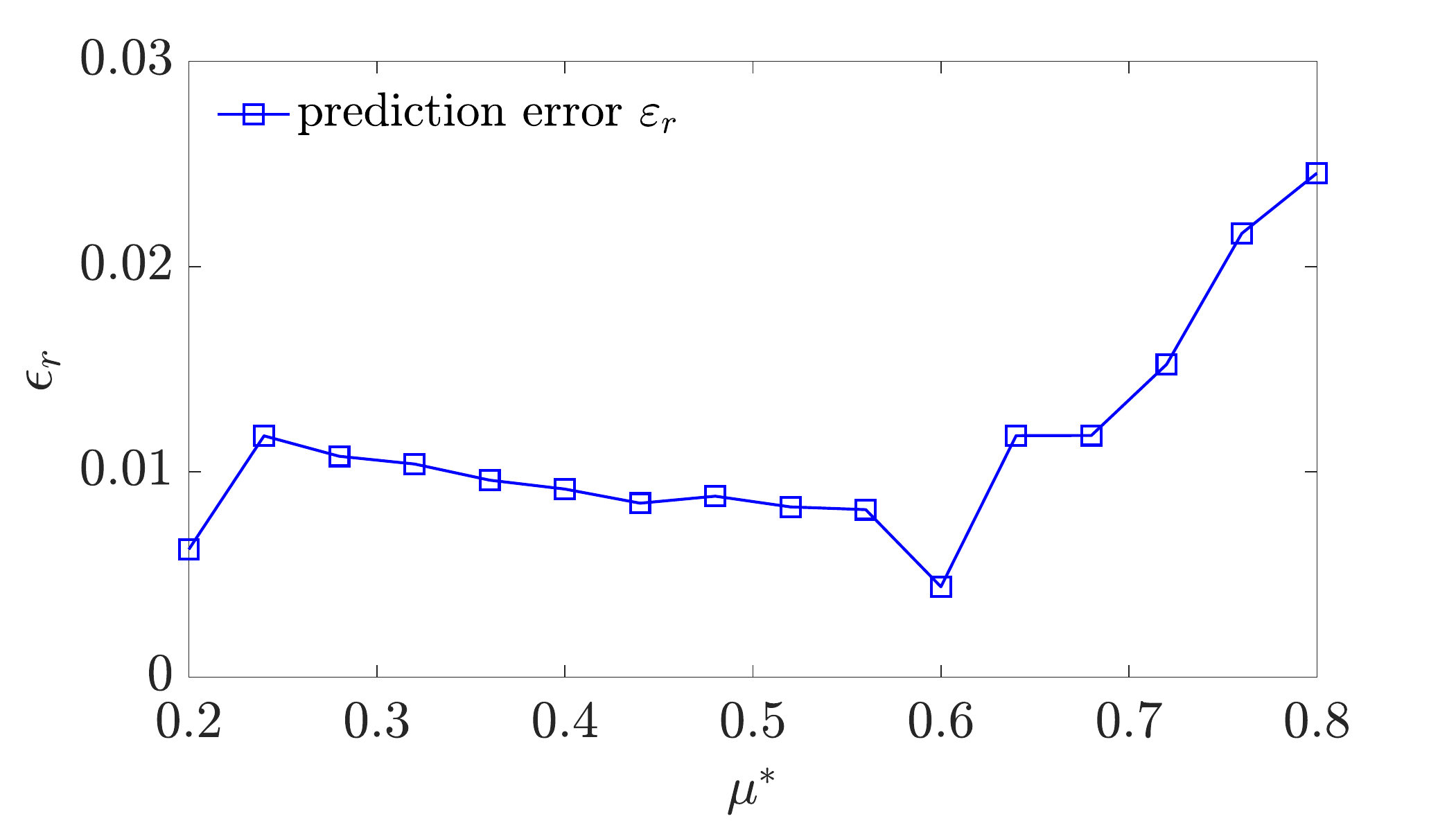}
    \caption{The relative error $\varepsilon_r(\mu^*)$ of the memory kernels $K^\text{R-G}(t;\mu^*)$ predicted by ROM-GPR transfer learning at different testing parameter instances $\mu^* \in \mathcal{P}_\text{test}$.}
    \label{fig:N31_error}
\end{figure}
To examine the accuracy of such predicted memory kernel, we selected a set of testing parameter instances: $\mathcal{P}_\text{test} = \{ 0.2+ 0.04i  \}_{i = 0,1,\dots,15}$. 
Note that while some of these testing parameter instances are within the range of training sampling space $\mathcal{P}_\text{tr}$, others are beyond the range; i.e., the predictions include not only interpolation but also extrapolation. We used the relative error to evaluate the accuracy of prediction, which is defined as: $\varepsilon_r = \frac{\| K^\text{R-G} - K  \|_2}{\| K  \|_2}$ with $K^\text{R-G}$ denoting the memory kernel predicted by ROM-GPR transfer learning, $K$ the memory kernel determined from MD simulation data, and $ \| \cdot \|_2 $ denoting $L_2$ norm. The relative errors of the memory kernels predicted at all 16 testing parameter instances are depicted in Fig.~\ref{fig:N31_error}. It can be seen that all predictions are accurate with the largest relative error of 2.5\%; extrapolation outside the training sampling space $\mathcal{P}_\text{tr}$ is less accurate than interpolation inside $\mathcal{P}_\text{tr}$; and the error increases as the extrapolation goes further.
 
\begin{figure}[htbp]
    \centering
    \includegraphics[width=8.8cm]{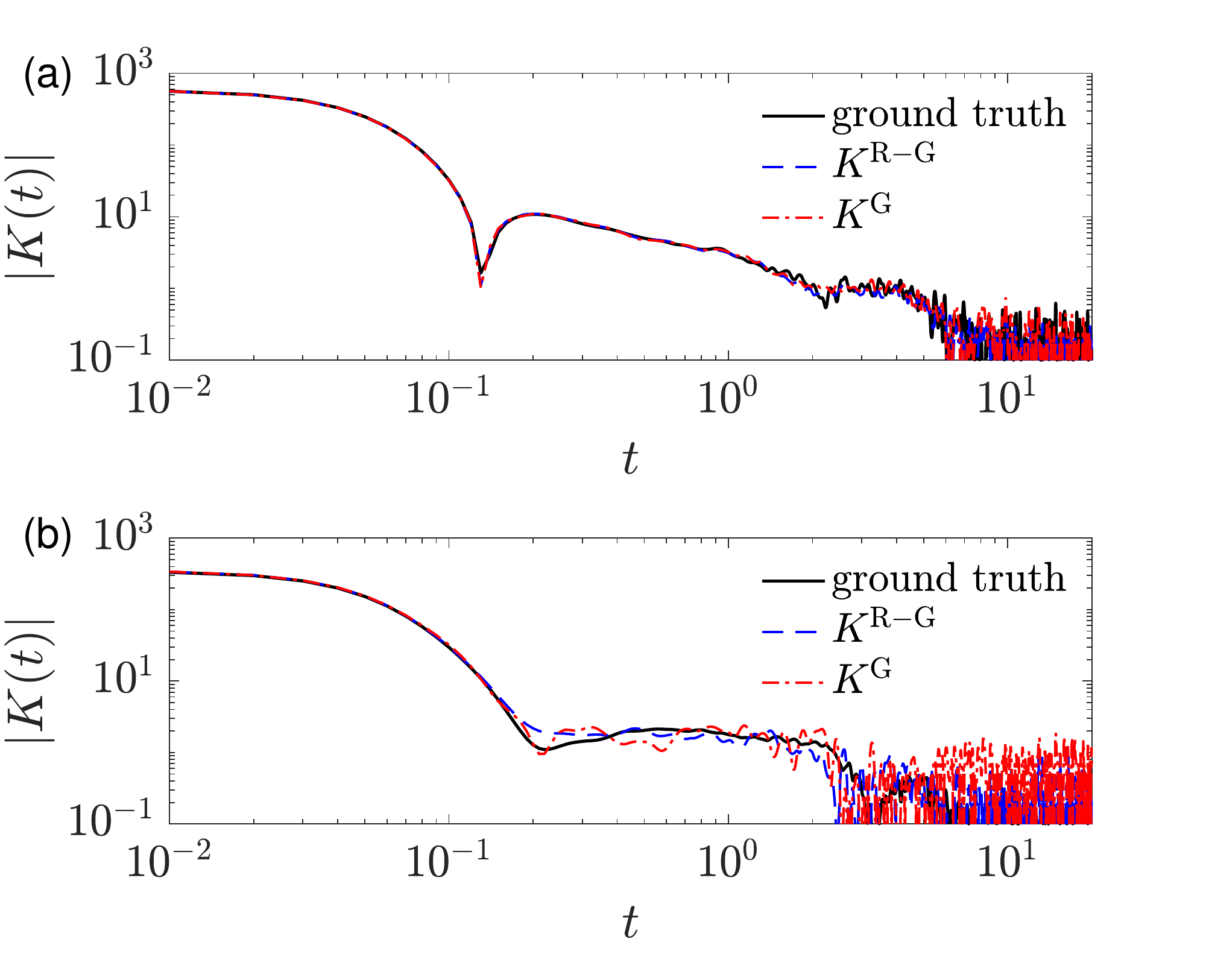}
    \caption{The memory kernels $K(t)$ predicted at (a) $\mu^*_1 = \gamma^*_1 = 0.4$ (interpolation) and (b) $\mu^*_2 = \gamma^*_2 = 0.8$ (extrapolation) by the ROM-GPR and GPR transfer learning methods, respectively, compared with the memory kernels computed from the MD simulation data (ground truth).}
    \label{fig:N31_pred1d}
\end{figure}

Next, we compare the ROM-GPR transfer learning method with the GPR transfer learning method in terms of the accuracy and computational cost. For that, we used the same training parameter instances determined by active learning, i.e.,  $\mathcal{P}^\text{active}_\text{tr} = \{ 0.2486, 0.4528, 0.6, 0.2 \}$; and two testing parameter instances were selected: $\mu^*_1 = 0.4$ inside the training sampling space $\mathcal{P}_\text{tr}$ and $\mu^*_2 = 0.8$ outside $\mathcal{P}_\text{tr}$. The memory kernels predicted at the two testing parameter instances by the two methods are compared in Fig.~\ref{fig:N31_pred1d}. By comparison with the memory kernel computed from the MD simulation data (regarded as the ``ground truth"), we find that the predictions of these two methods both achieve good accuracy with the relative errors $\varepsilon_r(\mu_1^*) = 0.009 $ and $\varepsilon_r(\mu_2^*) = 0.025$ for $K^\text{R-G}$ and 
$\varepsilon_r(\mu_1^*) = 0.011 $ and $\varepsilon_r(\mu_2^*) = 0.046 $ for $K^\text{G}$, where $K^\text{G}$ denotes the memory kernel predicted by the GPR transfer learning method.
However, the computational costs of the two methods are significantly different. Specifically, it took the GPR transfer learning method 1086s to construct the GPR model and 0.2832s to predict the memory kernel at a testing parameter instance, but only took the ROM-GPR transfer learning method 0.1127s to construct the ROM and GPR model and $7.880 \times 10^{-4}$s to predict $K(t;\mu^*)$. Both methods were implemented in {\sc Matlab} and run on a workstation with Intel Core CPU i5-6500 3.20 GHz. We hence evidence that the transfer learning based on GPR can be greatly accelerated by model order reduction. 

\begin{figure}[htbp]
    \centering
    \includegraphics[width=8.8cm]{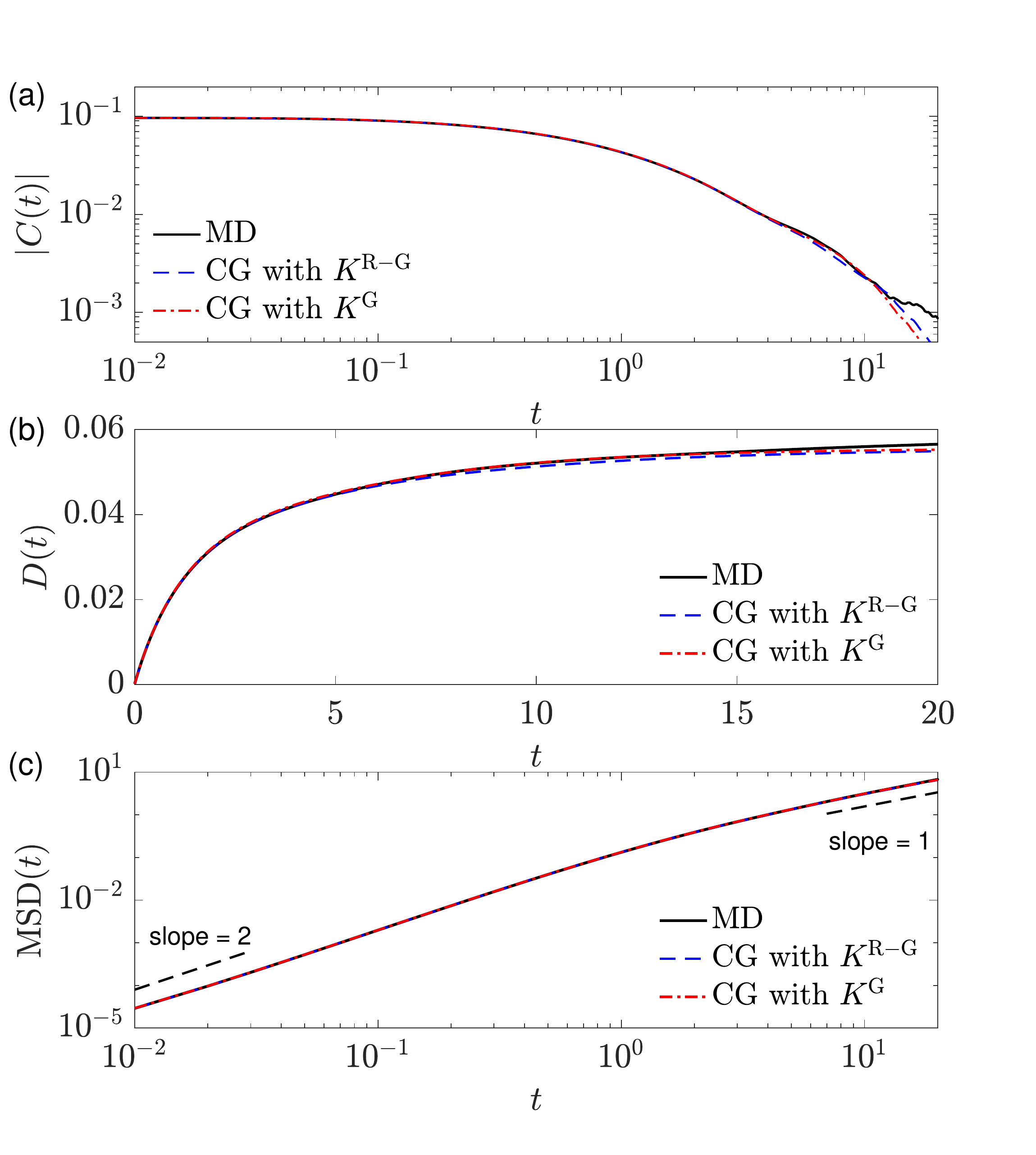}
    \caption{VACF ($C(t)$), diffusion coefficient ($D(t)$), and MSD predicted by the CG models at $\mu^*_1 = \gamma^*_1 = 0.4$, compared with the MD simulation results. Here, the memory kernel predicted by the ROM-GPR or GPR transfer learning method was approximated by Eq.~\eqref{equ:K_fit} with $\mathcal{N} = 7 $ terms.}
    \label{fig:N31_CG1d_mu1}
\end{figure}

\begin{figure}[htbp]
    \centering
    \includegraphics[width=8.8cm]{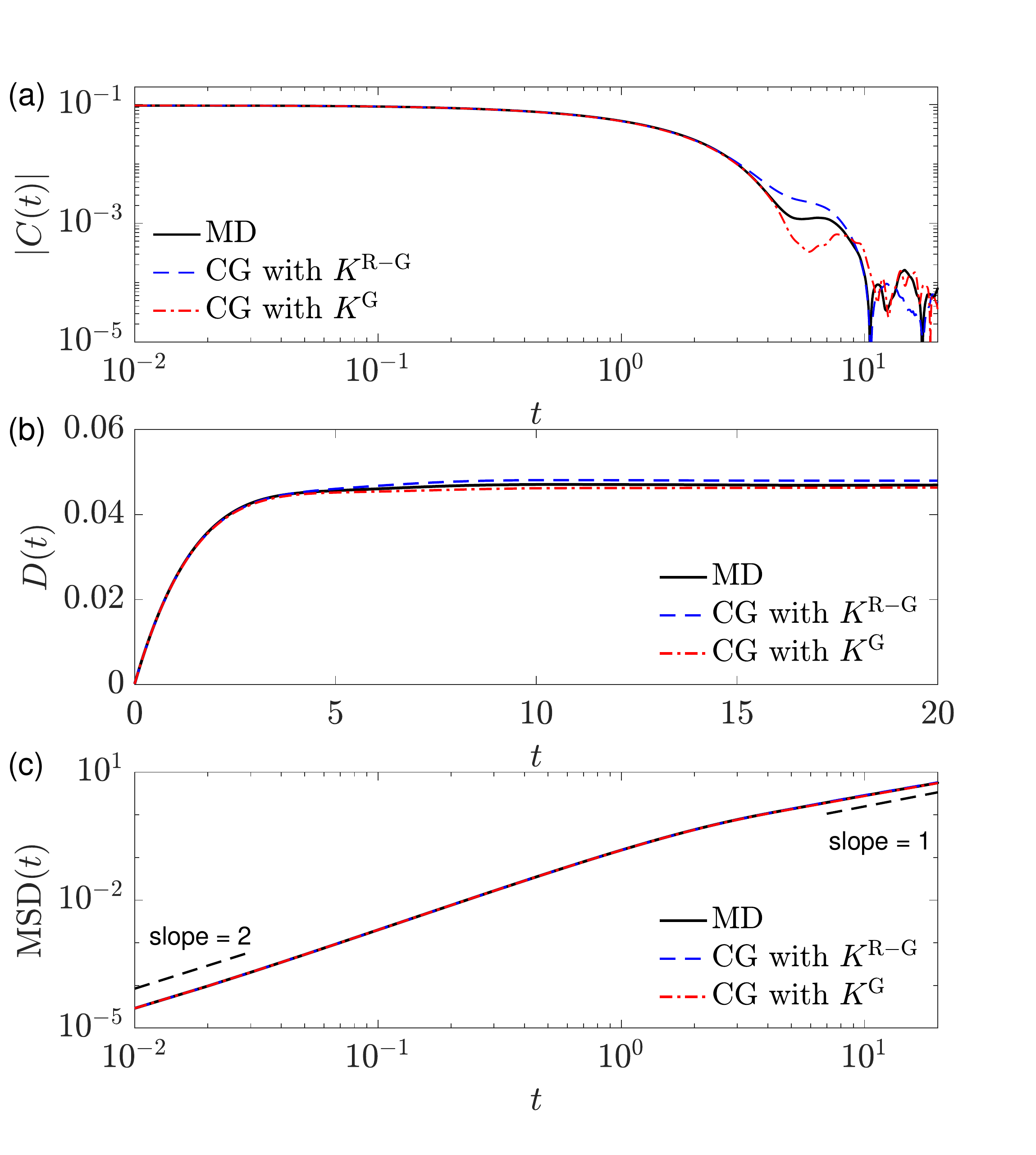}
    \caption{VACF ($C(t)$), diffusion coefficient ($D(t)$), and MSD predicted by the CG models at $\mu^*_2 = \gamma^*_2 = 0.8$, compared with the MD simulation results. Here, the memory kernel predicted by the ROM-GPR or GPR transfer learning method was approximated by Eq.~\eqref{equ:K_fit} with  $\mathcal{N} = 7 $ terms.} 
    \label{fig:N31_CG1d_mu2}
\end{figure}

Finally, from the predicted memory kernel at a given parameter instance, the CG model was constructed. For that, the memory kernel was first approximated by Eq.~\eqref{equ:K_fit}. The dynamics of the CG system was then simulated by solving Eq. \eqref{Eq:extend}. To examine whether the CG model could accurately reproduce the dynamic properties of the reference atomistic system in all time scales, we computed the VACF ($C(t)$), diffusion coefficient ($D(t)$), and MSD as functions of time. In Figs.~\ref{fig:N31_CG1d_mu1} and \ref{fig:N31_CG1d_mu2}, we present the results obtained from CG modeling at two different concentrations of star polymers ($\gamma^*_1 = 0.4$ and $\gamma^*_2 = 0.8$), compared with the MD simulation results. Overall good agreements can be found, and all stages of diffusion from super-diffusion to normal diffusion are correctly captured. We hence demonstrate that by transfer learning of 4 memory kernels at concentrations $\gamma=\{0.2486,0.4528,0.6,0.2\}$, we were able to construct the CG model that could accurately reproduce the reference atomistic system's dynamic properties in all time scales across a range of concentrations ($\gamma^* \in (0.2,0.8]$), unseen in the training data and even outside the range of training sampling space $\mathcal{P}_\text{tr} = [0.2,0.6]$. 

\subsubsection{Transferable in multiple parameters}
\label{subsubsec:N31-3d}
\begin{figure}[htbp]
    \centering
    \includegraphics[width=8.8cm]{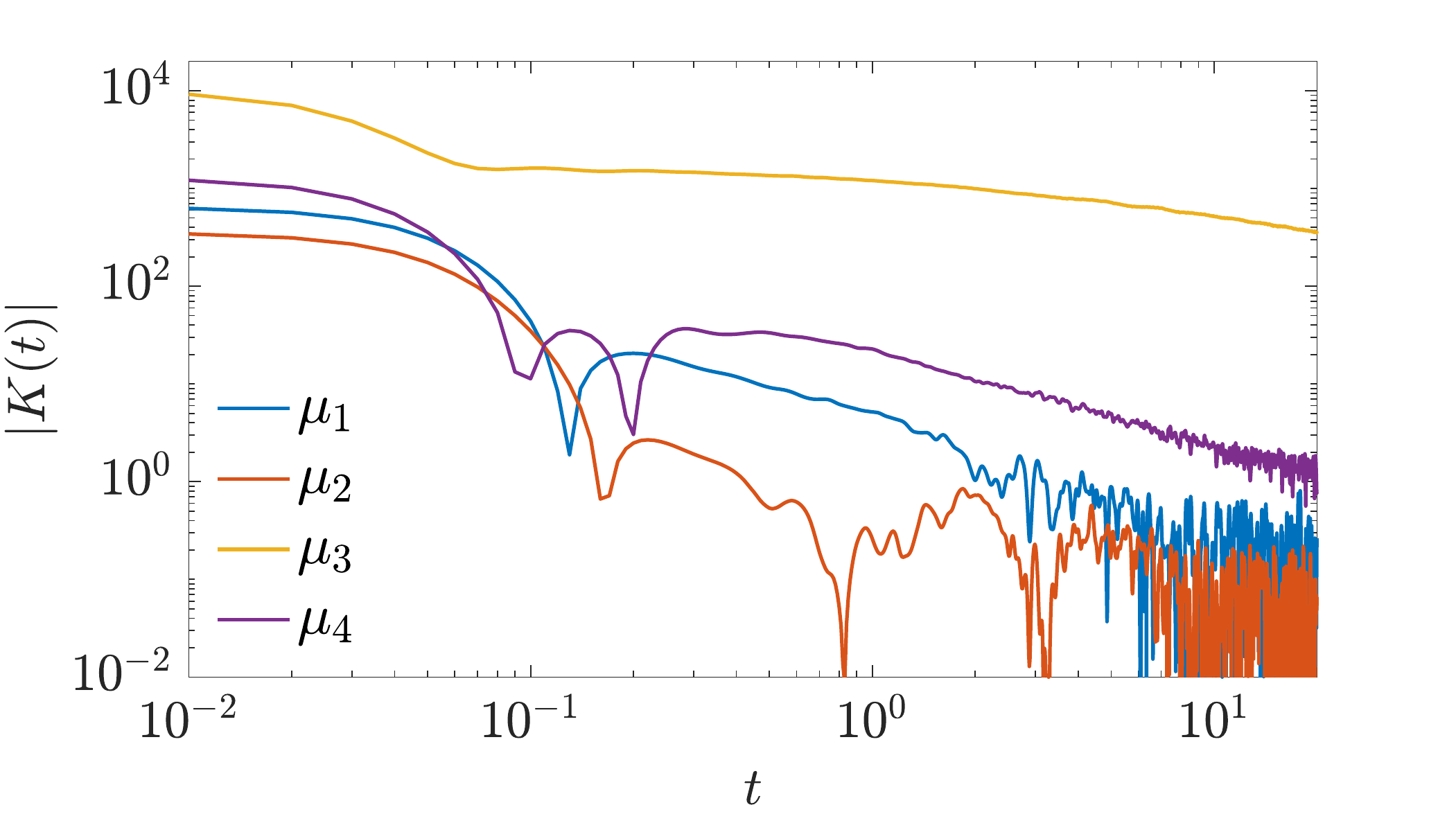}
    \caption{Sample memory kernels at 4 parameter instances: $\boldsymbol \mu_1 = (0.75,0.3,1.1)$, $\boldsymbol \mu_2 = (0.75,0.7,1.1)$, $\boldsymbol \mu_3 = (1.75,0.3,2.1)$, and $\boldsymbol \mu_4 = (1.75,0.7,2.1)$.}  
    \label{fig:N31_sample}
\end{figure}

After verifying the transferability of memory kernel in one parameter as well as a detailed discussion about the two transfer learning methods and the active learning strategy, we next demonstrate how transfer learning can  enable the memory kernel transferable in multiple parameters. In particular, the parameter space comprises three parameters: the temperature $T$, the concentration of star polymers $\gamma$, and the solvent viscosity characterized by $\sigma_\text{sol}$, and was specified as $\mathcal{P}= [0.5, 2.0] \times [0.2, 0.8] \times [1.0, 2.2]$ in this case with the training sampling space set as $\mathcal{P}_\text{tr} = [0.75,1.75] \times [0.3,0.7] \times [1.1,2.1]$. In such a multi-dimensional, large parameter space, the memory kernels' magnitudes and time scales vary in orders, as depicted in Fig.~\ref{fig:N31_sample}, where 4 memory kernel samples are displayed at 4 parameter instances sampled in $\mathcal{P}_\text{tr}$. This makes transfer learning challenging and hence serves as a good problem to assess the effectiveness of our proposed methods.

\begin{figure}[htbp]
    \centering
    \includegraphics[width=8.8cm]{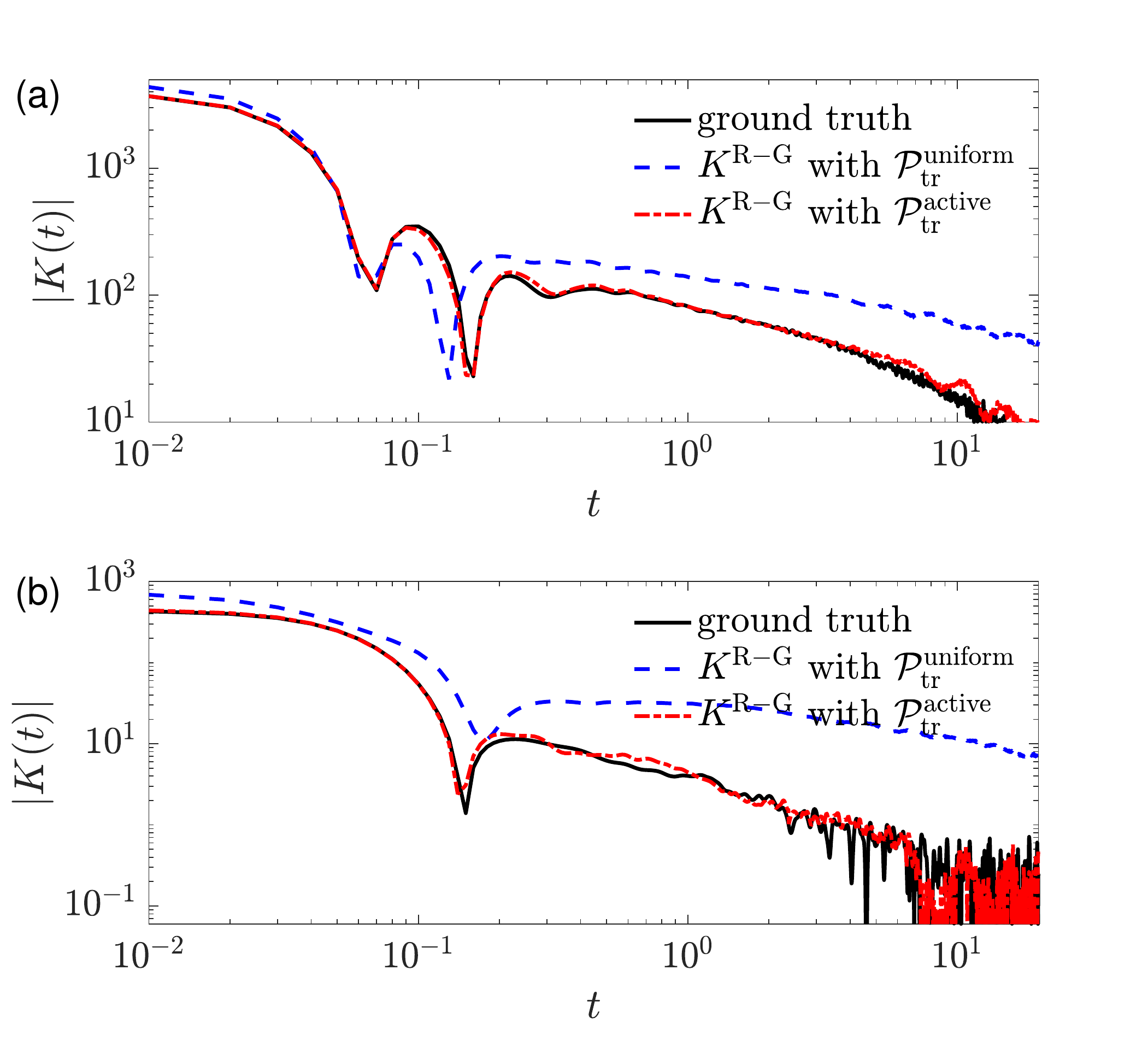}
    \caption{The memory kernels $K^\text{R-G}(t)$ predicted at (a) $\boldsymbol \mu^*_1 = (1.75,0.42,1.74) $ and (b) $\boldsymbol \mu^*_2 = (0.5,0.2,1.0)$ by the ROM-GPR transfer learning method with  $\mathcal{P}^\text{active}_{\text{tr}}$ vs. $\mathcal{P}^\text{uniform}_\text{tr}$, compared with the memory kernels directly computed from the MD simulation data (ground truth).}  
    \label{fig:N31_K_3d}
\end{figure}

The active learning process started with 8 initial training parameter instances chosen as $\mathcal{P}^\text{active}_{\text{tr}} = \{( 0.75 + 1i_1 , 0.3 + 0.4i_2, 1.1+1i_3 ) \}_{i_1 = 0,1;\ \ i_2 = 0,1; \ \ i_3 = 0,1 }$. For the preset tolerance $\zeta_\text{POD} = 0.01$, 8 POD bases were retained in the ROM, i.e., $R= 8 $ in Eq.~\eqref{Eq:ROM_POD}. As a result of active learning, $N_p = 54 $ sampling points were determined as the final training parameter instances, such that the uncertainty (defined in Eq.~\eqref{Eq:uncertainty}) of the GPR models for the parameter modes $\alpha_k(\boldsymbol{\mu})$ with $k=1,\dots,8$ was less than the preset tolerance $\zeta_\text{AL} =  0.025 $. To illustrate the advantage of active learning, we set up a control group, where the training parameter instances of the same number ($N_p = 54 $) were sampled \textit{uniformly} as: $\mathcal{P}^\text{uniform}_\text{tr} = \{( 0.75 + 0.5i_1 , 0.3 + 0.2i_2, 1.1+0.2i_3 ) \}_{i_1 = 0,1,2;\ \ i_2 = 0,1,2; \ \ i_3 = 0,\dots,5 }$. Two testing parameter instances were selected: $\boldsymbol \mu^*_1 = (1.75,0.42,1.74) $ inside the training sampling space $\mathcal{P}_\text{tr}$ but not in $\mathcal{P}^\text{active}_\text{tr}$ or $\mathcal{P}^\text{uniform}_\text{tr}$, and $\boldsymbol \mu^*_2 = (0.5,0.2,1.0) $ outside $\mathcal{P}_\text{tr}$. Fig.~\ref{fig:N31_K_3d} presents the memory kernels predicted by the ROM-GPR transfer learning method at the two testing parameter instances with different sets of training parameter instances, i.e., $\mathcal{P}^\text{active}_{\text{tr}}$ vs. $\mathcal{P}^\text{uniform}_\text{tr}$. By comparison with the ground truth, it is obvious that using the same number of training data, the memory kernels predicted using the training parameter instances adaptively sampled via active learning are more accurate than using the uniformly sampled training parameter instances. While the relative errors for the former are $\varepsilon_r(\mu_1^*) = 0.026 $ and $\varepsilon_r(\mu_2^*) = 0.068 $, the relative errors for the latter are $\varepsilon_r(\mu_1^*) = 0.349 $ and $\varepsilon_r(\mu_2^*) = 0.932 $. Thus, to achieve the same target accuracy, transfer learning coupled with active learning may require much less training data, which thereby greatly reduces the training cost.  

\begin{figure}[htbp]
    \centering
    \includegraphics[width=8.8cm]{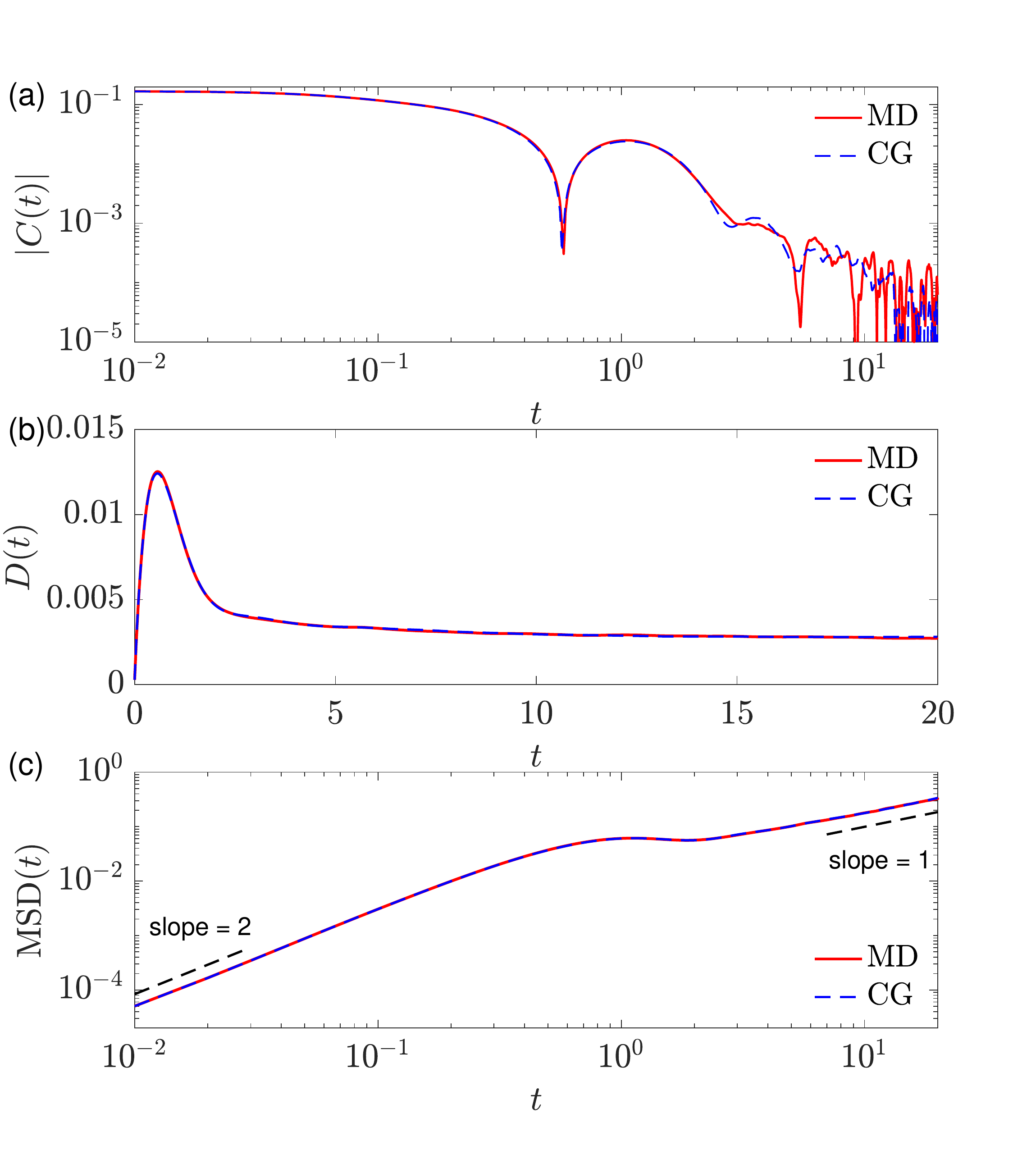}
    \caption{VACF ($C(t)$), diffusion coefficient ($D(t)$), and MSD predicted by the CG model at $\boldsymbol \mu^*_1 = (1.75,0.42,1.74)$, compared with the MD simulation results. Here, the memory kernel predicted by the ROM-GPR transfer learning method coupled with active learning was approximated by Eq.~\eqref{equ:K_fit} with $\mathcal{N} = 5 $ terms.}
    \label{fig:N31_CG3d_mu1}
\end{figure}

\begin{figure}[htbp]
    \centering
    \includegraphics[width=8.8cm]{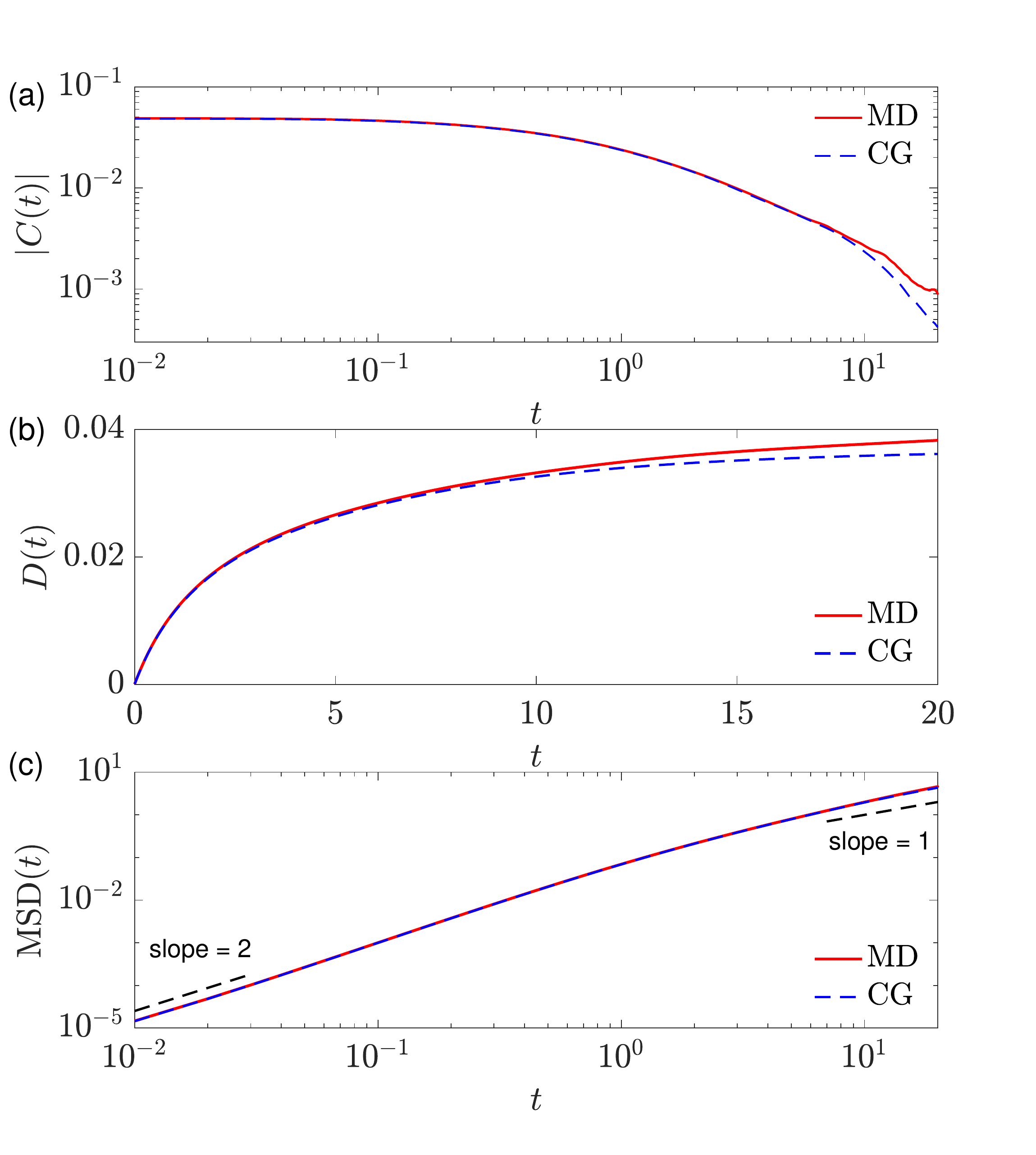}
    \caption{VACF ($C(t)$), diffusion coefficient ($D(t)$), and MSD predicted by the CG model at $\boldsymbol \mu^*_2 = (0.5,0.2,1.0)$, compared with the MD simulation results. Here, the memory kernel predicted by the ROM-GPR transfer learning method coupled with active learning was approximated by Eq.~\eqref{equ:K_fit} with $\mathcal{N} = 6 $ terms.}
    \label{fig:N31_CG3d_mu2}
\end{figure}

The CG model was then constructed from the predicted $K^\text{R-G}(t)$. Figs.~\ref{fig:N31_CG3d_mu1} and \ref{fig:N31_CG3d_mu2} present the VACF, diffusion coefficient, and MSD predicted by the CG model. They agree well with the MD simulation results. All stages of diffusion: super-, sub-, and normal diffusion (in Fig.~\ref{fig:N31_CG3d_mu1}) are correctly captured. We hence verify that by transfer learning, the memory kernel is transferable in multiple parameters, built on which the CG models can accurately reproduce the dynamic properties of the reference atomistic system in all time scales for different temperatures, concentrations of star polymers, and solvent viscosities. 


\subsection{Peptoid polymer}
\begin{figure}[htbp]
    \centering
    \includegraphics[width=5cm]{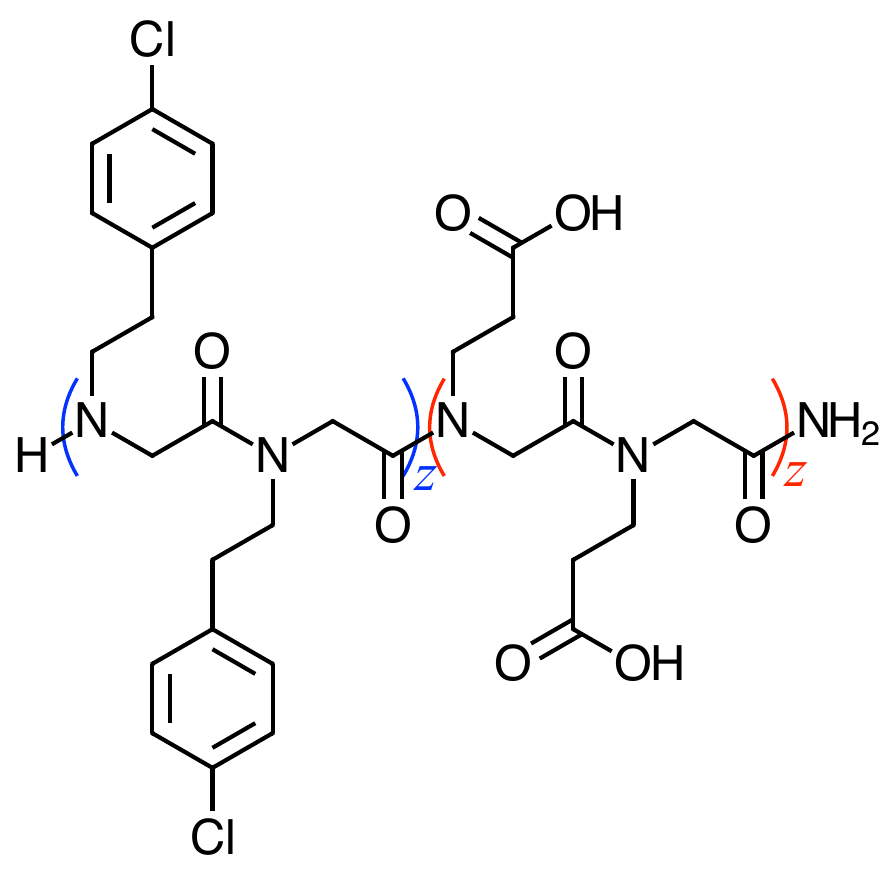}
    \caption{Chemical structure of a peptoid polymer. $z$ denotes the number of repeat units: $\mathrm{N_{4-Cl}pe}$ and Nce.}  
    \label{fig:peptoid_stru}
\end{figure}

\begin{figure}[htbp]
    \centering
    \includegraphics[width=7cm]{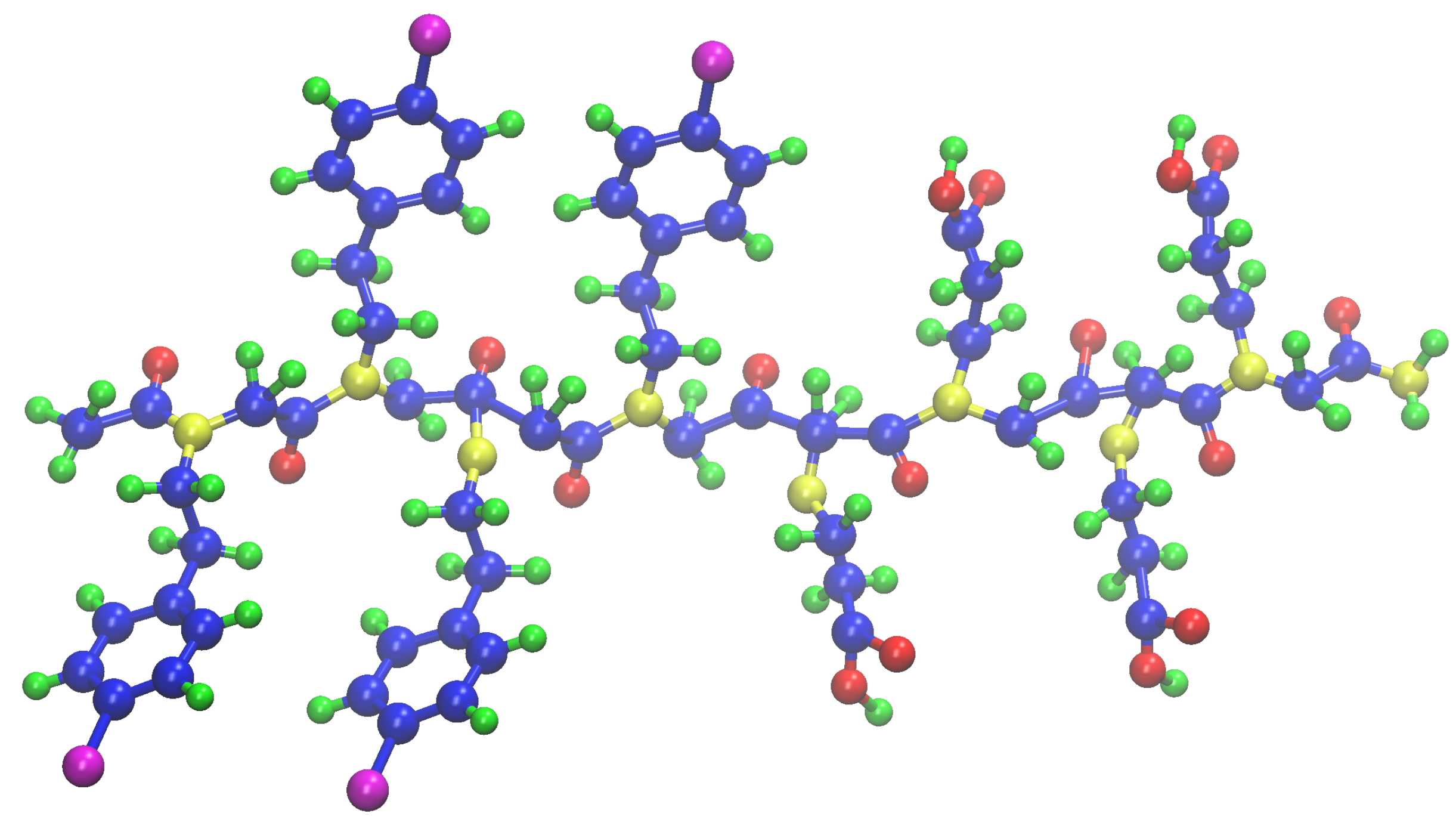}
    \caption{Atomistic representation of a peptoid polymer with the number of repeat units $z=2$.}  
    \label{fig:peptoid_atomistic}
\end{figure}

After assessing the proposed methods in a model polymer solution system, we next studied a real polymer system consisting of peptoid polymers, which are known as a class of highly configurable biopolymers \cite{Peptoid2_2013,jin2016highly,PeptoidReview_CR2016,Peptoid_ChenNM2017,Peptoid_Application2019,Peptoid_RonSM2019,Peptoid1_2021}. The peptoid polymer studied in this work contains N-[2-(4-chlorophenyl)ethyl]glycine ($\mathrm{N_{4-Cl}pe}$) and  N-(2-carboxyethyl)glycine (Nce) groups \cite{jin2016highly}, whose chemical structure is described in Fig.~\ref{fig:peptoid_stru}. In different solution systems, a peptoid polymer can consist of different numbers of repeat units, denoted as $z$. The peptoid solutions were made of one peptoid polymer immersed in water with a fixed concentration of 1600 mg/kg. The all-atom representation of a peptoid polymer in MD simulations is shown in Fig.~\ref{fig:peptoid_atomistic}. Here, the AMBER03~\cite{duan2003point} force field with the parameters from the generalized AMBER force field~\cite{wang2006automatic,wang2004development,jin2016highly} was adopted for the peptoid molecule; the SPC force field~\cite{berendsen1981intermolecular} was employed for modeling the water molecules. The MD simulations were performed using the GROMACS package~\cite{abraham2015gromacs}. 

The appropriate size of the periodic cubic box was determined by running the MD simulation in the isothermal–isobaric ensemble (NPT) using the Parrinello-Rahman barostat \cite{parrinello1981polymorphic} at 1 bar with isotropic pressure coupling. 
The MD simulations were then performed in the NVT ensemble with the modified Berendsen thermostat~\cite{bussi2009isothermal}. To maintain the entire system in the NVT ensemble more effectively, the thermostat was enforced to the peptoid molecule and to the water molecules separately. The time step $\Delta t$ was chosen as 2 fs. After the thermal equilibrium was reached (after 100 ps in our case), 
data were collected for computing the ensemble-averaged quantities of interest.

In this study, the temperature and the length (i.e., the number of repeat units) of a peptoid polymer varied in different systems. Hence, they were the two parameters to practice transfer learning and to assess the transferability of memory kernel. The parameter space was specified as $\mathcal{P} = [275, 365] \times [1,6]$, 
i.e., the temperature $T$ varies from 275 K to 365 K, and the number of repeat units in a peptoid polymer $z$ varies from 1 to 6. The training sampling space was set as $\mathcal{P}_\text{tr}  = [275, 315] \times [1,3]$. The 4 initial training parameter instances for active learning were sampled as $\mathcal{P}^\text{active}_\text{tr} = \{ (275 + 40i_1, 1+2i_2) \}_{i_1 = 0,1;\ \ i_2 = 0,1}$. The tolerance for POD was set as $\zeta_\text{POD} = 0.1$, for which 2 POD bases were retained in the constructed ROM with $R = 2 $ in Eq.~\eqref{Eq:ROM_POD}. After adding 6 more sampling points via active learning, the uncertainty (defined in Eq.~\eqref{Eq:uncertainty}) of the GPR models for the parameter modes $\alpha_k(\boldsymbol{\mu})$ with $k=1,2$ was less than the preset tolerance $\zeta_\text{AL} =  0.01 $. Thus, we used the memory kernels computed from the MD simulation data at the 10 sampled parameter instances as the training data in the transfer learning. To ensure the memory kernel long enough to capture the dynamics in all time scales, each was computed up to $t_f = 50$ ps until the VACF ($C(t_f)$) decayed to $ |C(t_f) |/|C(0) |\leq 10^{-2} $.
\begin{figure}[htbp]
    \centering
    \includegraphics[width=8.8cm]{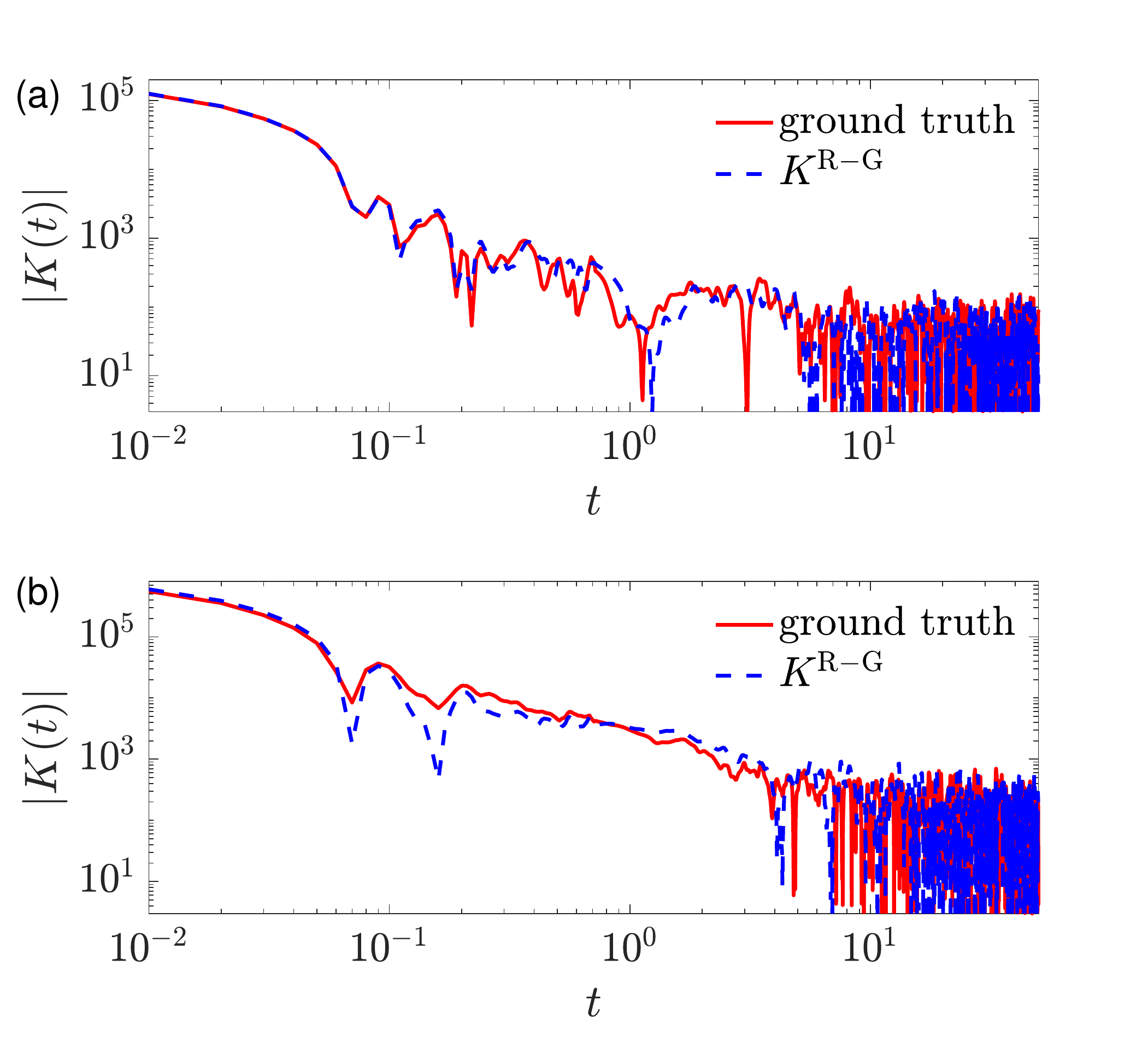}
    \caption{The memory kernels $K^\text{R-G}(t)$ predicted by the ROM-GPR transfer learning method at (a) $\boldsymbol \mu^*_1 = (307,1) $ (interpolation), corresponding to $T=$307 K and $z=1$ and (b) $\boldsymbol \mu^*_2 = (365,6)$ (extrapolation), corresponding to $T=$365 K and $z=6$, compared with the memory kernels directly computed from the MD simulation data (ground truth).}
    \label{fig:peptpoid_K_pred}
\end{figure}

To examine the performance of transfer learning, two testing parameter instances were selected: $\boldsymbol \mu^*_1 = (307,1)$, which is where the uncertainty of the GPR models for all parameter modes
is maximized inside the training sampling space $\mathcal{P}_\text{tr}$; and $\boldsymbol \mu^*_2 = (365,6)$, which is outside $\mathcal{P}_\text{tr}$ and represents the most extreme case in the parameter space considered. Fig.~\ref{fig:peptpoid_K_pred} shows the memory kernels predicted at the two selected testing parameter instances. It can be seen that the memory kernels predicted by the ROM-GPR transfer learning agree well with the ground truths with the relative errors of 0.028 and 0.086 for $\boldsymbol \mu^*_1$ and $\boldsymbol \mu^*_2$, respectively.

\begin{figure}[htbp]
    \centering
    \includegraphics[width=8.8cm]{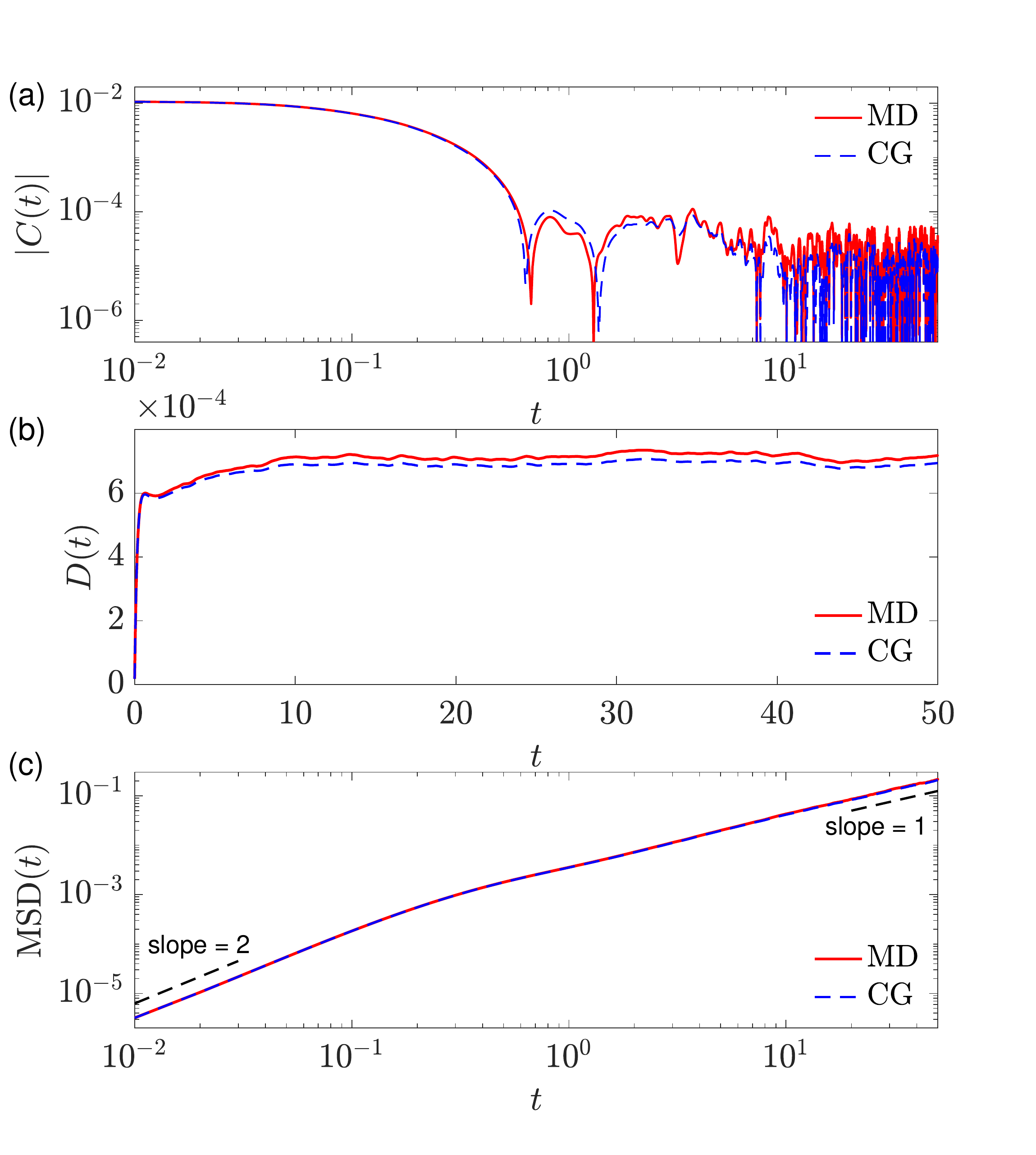}
    \caption{VACF ($C(t)$), diffusion coefficient ($D(t)$), and MSD predicted by the CG model at $T=$307 K for the peptoid polymer with one $\mathrm{N_{4-Cl}pe}$ and Nce, compared with the MD simulation results. Here, $K^\text{R-G}(t; \boldsymbol \mu^*_1)$ was approximated by Eq.~\eqref{equ:K_fit} with $\mathcal{N} = 7 $ terms.}
    \label{fig:peptoid_CG_mu1}
\end{figure}

\begin{figure}[htbp]
    \centering
    \includegraphics[width=8.8cm]{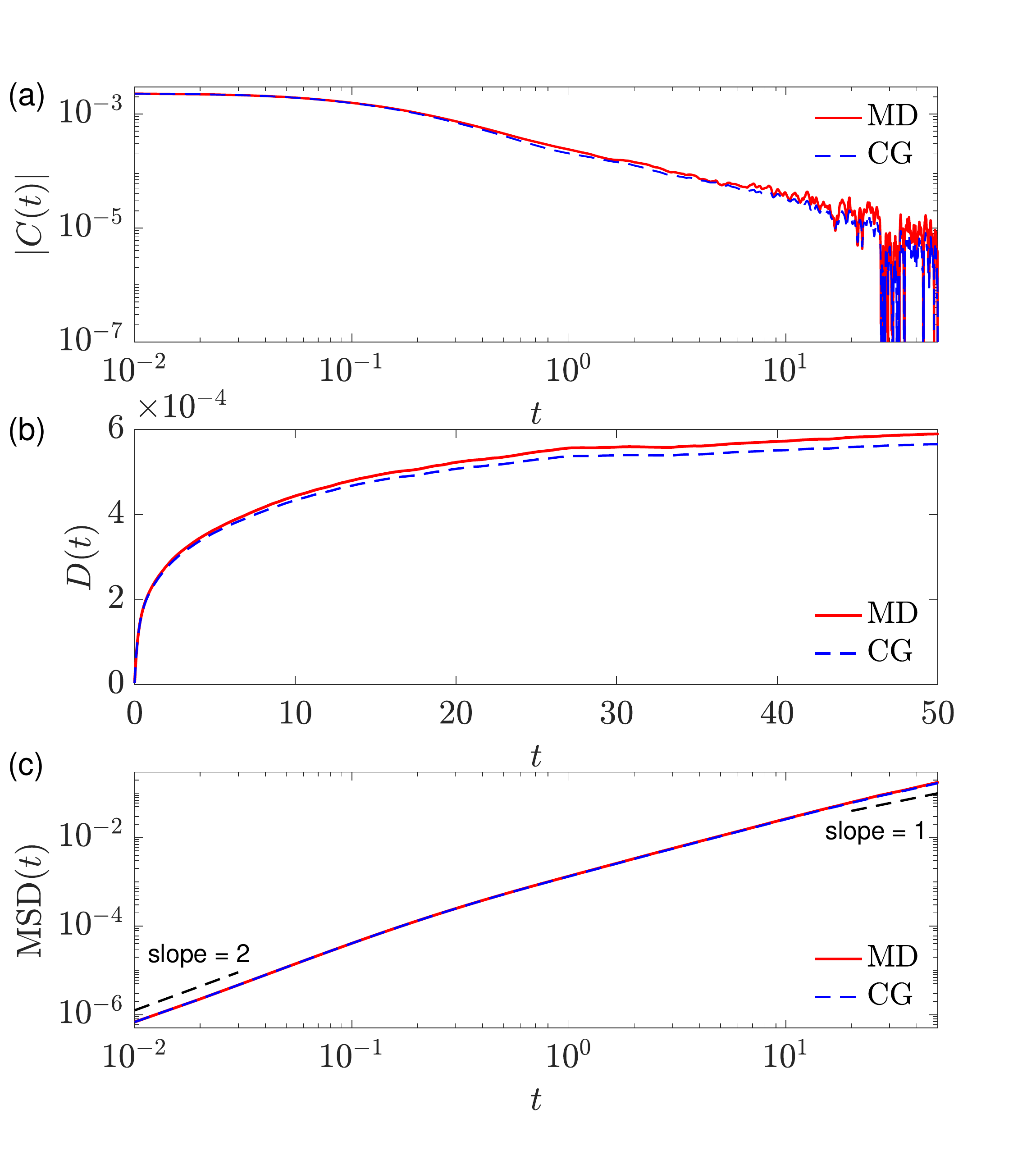}
    \caption{VACF ($C(t)$), diffusion coefficient ($D(t)$), and MSD predicted by the CG model at $T=$365 K for the peptoid polymer with 6 $\mathrm{N_{4-Cl}pe}$ and Nce, compared with the MD simulation results. Here, $K^\text{R-G}(t; \boldsymbol \mu^*_2)$ was approximated by Eq.~\eqref{equ:K_fit} with $\mathcal{N} = 8 $ terms.}
    \label{fig:peptoid_CG_mu2}
\end{figure}

Using the predicted memory kernels $K^\text{R-G}(t; \boldsymbol \mu^*)$, we constructed the CG models. The VACF, diffusion coefficient, and MSD predicted by the CG models were compared with the MD simulation results, as depicted in Figs.~\ref{fig:peptoid_CG_mu1} and \ref{fig:peptoid_CG_mu2}. Good agreements were achieved in all time scales. Thus, in a real polymer solution system, we again verify that by transfer learning, the memory kernel can be transferable in multiple parameters and across a non-trivially large parameter space. Built on the transferable memory kernel, the CG model can accurately reproduce the dynamic properties of the peptoid polymer in all time scales for different temperatures and lengths of polymer.

\section{Conclusion}\label{sec:conclusion}
We have introduced transfer learning of memory kernel in CG modeling. Built on a transferable memory kernel, the CG model can reproduce the dynamic properties of the reference atomistic system at different thermodynamic conditions such as temperature and solvent viscosity and for different systems with varying solute concentrations and lengths of polymer chains. Notably, the transferability allows for out-of-sample predictions in the extrapolated domain of parameters. 
Through two example polymer solution systems, we have demonstrated the accuracy and efficiency of the proposed transfer learning.
Since the memory kernels are accurately predicted, the CG models can reproduce the dynamic properties of polymers in all time scales, including the VACF, diffusion coefficient, and MSD. The predictions of CG models agree well with the atomistic simulation results, and all stages of diffusion: super-, sub-, and normal diffusion, are correctly captured. We anticipate the proposed methodology is generally applicable to CG modeling of other kinds of polymer solutions and soft matter systems such as colloid suspensions. While the parameters considered in this work include the temperature, solvent viscosity, concentration of polymers, and length of polymer chains, the proposed transfer learning can certainly be applied to a broader range of parameters beyond this list.  

Two transfer learning methods have been proposed and compared. The GPR transfer learning method directly models memory kernels (functions of time) at different parameters as a multivariate Gaussian process. GPR is not only flexible for interpolation or extrapolation at any given input, but also can quantify the uncertainty of any prediction. The ROM-GPR transfer learning method integrates the GPR with model order reduction and active learning, by which it is more computationally efficient and requires minimum training data. In the ROM-GPR transfer learning method, a ROM is first constructed via POD for the memory kernel. By such, the memory kernel is represented in a reduced temporal and parameter space, and the GPR is only needed for the parameter modes. Thus, both training and prediction costs of GPR are greatly reduced. Furthermore, guided by the uncertainty quantified in GPR, the active learning technique enables adaptively sampling the training data. Compared with other sampling strategy, e.g., uniform sampling, using the same number of training data, active learning leads to much more accurate transfer learning. Thus, to achieve the same accuracy, active learning requires less training data and thereby lower training cost.

The present work has attempted to construct transferable memory kernels that preserve dynamic properties under coarse-graining. It can be potentially integrated with the efforts in literature that focus on attaining transferable CG potentials  \cite{CGGNN_JCP2020,CGGNN_Ionic_JCP2020,CGTrans_JCTC2020,CGTrans_JCTC2017,CGTrans_JCP2020}, so that the CG modeling can preserve both structural and dynamic properties of the underlying atomistic system and is transferable across a range of parameters.

\section*{Conflicts of interest}
There are no conflicts to declare.

\section*{Acknowledgements}
Z. M., W. P., M. K., and K. L. acknowledge the funding support from the Defense Established Program to Stimulate Competitive Research (DEPSCoR) Grant No. FA9550-20-1-0072. S. W. and W. P. acknowledge the funding support from the National Science Foundation under Grant No. CMMI-1761068. Pacific Northwest National Laboratory (PNNL) is multiprogram national laboratory operated for Department of Energy by Battelle under Contracts No. DE-AC05-76RL01830.

\balance

\bibliography{rsc1,rsc2,rsc3} 
\bibliographystyle{rsc} 

\end{document}